\documentclass[sigconf, nonacm=True]{acmart}
\usepackage{color}
\newcommand{\subheader}[1]{\vspace{0.1cm}\noindent\textbf{#1}.}

\usepackage[normalem]{ulem}

\usepackage{bm}
\usepackage{inputenc}
\usepackage{amsmath}
\usepackage{subcaption}

\setcopyright{none} 
\usepackage{comment}
\usepackage{appendix}

\mathchardef\UrlBreakPenalty=32767

\begin{document}

\title[Simulating systematic bias in attributed social networks]{Simulating systematic bias in attributed social networks\newline and its effect on rankings of minority nodes}

\author{Felix I. Stamm}
\authornote{Both authors contributed equally to this research.}
\email{felix.stamm@cssh.rwth-aachen.de}
\affiliation{%
  \institution{RWTH Aachen University}
  \country{Germany}
}

\author{Leonie Neuhäuser}
\authornotemark[1]
\email{neuhaeuser@cs.rwth-aachen.de}
\affiliation{%
  \institution{RWTH Aachen University}
  \country{Germany}
}

\author{Florian Lemmerich}
\email{florian.lemmerich@cssh.rwth-aachen.de}
\affiliation{%
  \institution{RWTH Aachen University}
  \country{Germany}
}

\author{Michael T. Schaub}
\email{schaub@cs.rwth-aachen.de}
\affiliation{%
  \institution{RWTH Aachen University}
  \country{Germany}
}

\author{Markus Strohmaier}
\email{markus.strohmaier@cssh.rwth-aachen.de}
\affiliation{%
  \institution{RWTH Aachen University \& GESIS}
  \country{Germany}
}

\renewcommand{\shortauthors}{Stamm, Neuhäuser}

\begin{abstract}
    Network analysis provides powerful tools to learn about a variety of social systems.
    However, most analyses implicitly assume that the considered relational data is error-free, reliable and accurately reflects the system to be analysed.
    Especially if the network consists of multiple groups, this assumption conflicts with a range of systematic biases, measurement errors and other inaccuracies that are well documented in the literature.
    To investigate the effects of such errors we introduce a framework for simulating systematic bias in attributed networks.
    Our framework enables us to model erroneous edge observations that are driven by external node attributes or errors arising from the (hidden) network structure itself.
    We exemplify how systematic inaccuracies distort conclusions drawn from network analyses on the network analysis task of minority representations in degree-based rankings.
    By analysing synthetic and real networks with varying homophily levels and group sizes, we find that introducing systematic edge errors can result both in a strongly increased or decreased ranking of the minority.
    The observed effect depends both on the type of edge error considered and level of homophily in the system. 
    We thus conclude that the implications of systematic bias in edge data depend on an interplay between network topology and type of systematic error.
    This emphasises the need for an error model framework as developed here, which provides a first step towards studying the effects of systematic edge-uncertainty for various network analysis tasks.
\end{abstract}

\def\UrlFont{\rmfamily\small}
\mathchardef\UrlBreakPenalty=10000

\maketitle

\section{Introduction}
\label{sec:Introduction}
\begin{figure*}[ht!]
\centering
\begin{subfigure}{0.24\linewidth}
\includegraphics[]{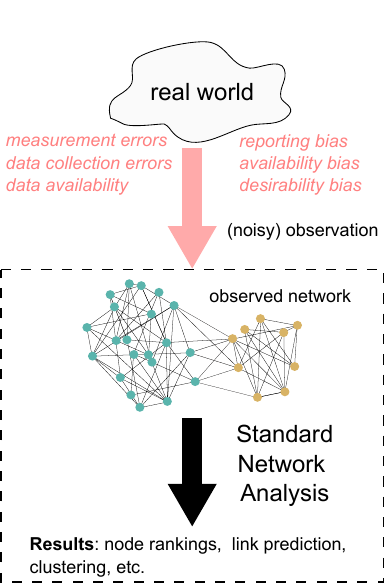}
\caption{ typical network analysis }
\end{subfigure}
\begin{subfigure}{0.74\linewidth}
\includegraphics[]{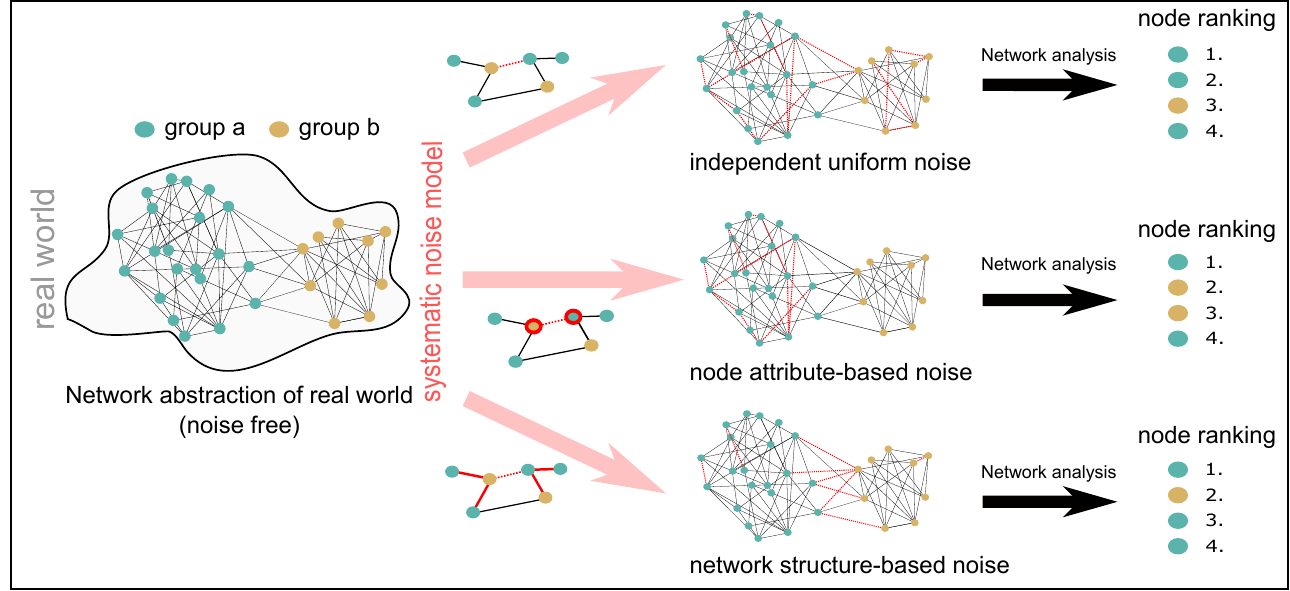}
\caption{ approach taken in this work }
\end{subfigure}
\caption{\textit{Social network analysis (a) without and (b) with considering systematic noise.} { (a) A typical approach to social network analysis that implicitly assumes an observed network without noise. (b) Our approach allows to simulate different kinds of edge noise, which we can compare to the assumed noise-free scenario.
This allows to quantify how noise can affect subsequent network analysis. In this paper we use node rankings based on degree centrality as example network analysis task. For more details of the node attribute based noise see Fig.~\ref{fig:metadatanoise}.}}
\label{fig:concept}
\end{figure*}

\newcommand{\mts}[1]{\textcolor{blue}{MS: #1}}
Networks have proven to be a simple but powerful model for the complex interplay of elements in systems from various areas, including nature, society and technology \cite{strogatz_exploring_2001,easley_networks_2010,newman_networks_2018}.
Computational network analysis typically processes observed networks as if these networks were free of uncertainties and accurately represent the system to be analysed.
While researchers have investigated how network analysis tasks are affected by potentially inaccurate observation of the node set \cite{wagner_sampling_2017,kossinets_effects_2006,borgatti_robustness_2006} and examined the consequences of random erroneous observation of edges~\cite{borgatti_robustness_2006,frantz_robustness_2009,lee_statistical_2006,murai_estimating_2019,wang_measurement_2012,almquist_random_2012}, non uniform and systematic erroneous edge observations~\cite{adiga_how_2013} have been largely neglected.
This seems in contrast with the large body of literature on various reporting biases, data collection biases, and other systematic errors which arise when data on social systems is collected~\cite{sen2019total,holland_structural_1973, feld_detecting_2002,brewer_forgetting_2000,marsden_network_1990,marsden_interviewer_2003}.
In this work we thus provide a model to study the impact of systematically missing edges on subsequent network analysis tasks.

Whether collected via surveys or through automated processes such as crawling, the vast majority of social networks describe connections that are reported or initiated by people and are thus unlikely to be completely bias free.
This presence of biases in the observation or reporting of network edges can lead to erroneous conclusions drawn from subsequent analysis, which becomes particularly problematic if results of the network analysis are used to drive real-world decisions, e.g. rankings of people in recruiting, marketing or other contexts. 
In social networks, edge errors may specifically depend on the presence of different social groups, which can be modeled as node labels in attributed networks~\cite{newman_structure_2016,yang_community_2013,wagner_sampling_2017} . 
Understanding how systematic edge noise may alter the results of network analyses is thus important from at least two perspectives: (i) from a scientific point of view we need to reach robust conclusions about the (social) systems we study; (ii) from a societal point of view we want to ensure, e.g., that our network analysis does not discriminate against specific groups.
Indeed, previous studies have shown that minorities can be disadvantaged in rankings~\cite{karimi_homophily_2018} due to network effects.

\subheader{Research Question}
We thus ask: how can systematic edge errors in attributed networks be simulated, what are its effects on subsequent network analysis, and how can such effects be assessed?

To illustrate these effects we focus on the question of minority representations in terms of degree-based node rankings, as a simple, but important prototypical network centrality analysis task. 
Such rankings based on the degree-centrality of the nodes are frequently used to access the relevance and importance of entities in networks.  

\subheader{Contributions and Findings}
(i) We introduce a general model for simulating systematic edge errors in attributed networks.
In this model we differentiate between (1) node attribute-based errors which are a function of the labels associated to the nodes an edge is incident on and (2) network structure-based errors which are independent of the node meta data and derive from the structure of the latent social network.
To illustrate how this error model can be used to assess the potential impact of systematic edge errors on network analysis, we focus on the representation of minorities in centrality-based rankings. 
We consider this a first step to study the effects of systematic noise, as simulated here by our error model, as centrality measures are frequently used to access the relevance of entities in social networks. Therefore, an altered representation of minorites in centrality-based rankings, caused by systematic noise in data, serves well as an indicator for misleading real-world conclusions.

(ii) We apply our method to synthetic and empirical networks with binary node labels. We find that systematic noise affects minority representation in rankings while uniform noise does not. 
The effect depends on both the type of edge error and the group-label assortativity in the network. 
While our modeling approach is general and can be easily extended to networks with multiple groups represented through nominal attributes, or to continuous node-covariates, we restrict our analysis here to the case of binary attributes representing two groups as an important but comparably simple setup.

\subheader{Implications}
With our model we enable researchers to simulate errors due to different data collection and reporting biases. 
As our numerical experiments show, the effects of such systematic errors on subsequent network analyses can vary depending on the network structure.
This emphasizes the utility of a framework such as ours, which enables researchers to simulate systematic error hypothesis and examine the effects on particular data sets.
While it may seem desirable to not rely on some hypothesis about the errors present in the data, we remark that unless strong assumptions are made about the generative process creating the observed network~\cite{peixoto_reconstructing_2018,newman_network_2018,young_robust_2020}, it is generally impossible to extract (and correct) errors in network data based on a single network observation.

\subsection*{Background and related work}
\label{sec:background}

\subheader{Networks with edge uncertainty}
Most empirical network analyses assume that the collected data accurately represent the underlying social system, even though such network data are known in general to be inherently noisy due to various social, cognitive and technical biases and errors. 
Especially if these network representations are used in subsequent data analysis tasks with real-world implications---consider, e.g., recommendation systems, importance rankings via centrality, etc.---neglecting such errors can have potentially detrimental consequences.
Focusing on the issue of edge-errors, the literature related to this work may be divided into two strands.

First, for \emph{unlabeled} networks a number of studies, e.g., ~\cite{borgatti_robustness_2006,frantz_robustness_2009,lee_statistical_2006,murai_estimating_2019,wang_measurement_2012,almquist_random_2012} have investigated the effect of random (non systematic) errors on certain analysis tasks. Some recent work has focused on inferring network errors, which is however only possible if the exact form of the network model and the error (unbiased, random) is assumed \cite{peixoto_reconstructing_2018,newman_network_2018,young_robust_2020,guimera_missing_2009}.
Similar to these studies we consider that there exists an underlying network, from which we can only obtain a noisy observation.
While there are some initial attempts at exploring non-uniform edge error models~\cite{adiga_how_2013}, we provide here a more complete characterization of errors and associated interpretations.
In addition, our error model allows for attribute dependent noise, which appears to have not been considered before in the literature.
We argue that systematic errors that are dependent on node attributes such as group membership are common and that attributed networks provide a unique opportunity to study the impact of these systematic uncertainties.

Second, noise in network data has been considered in the context of the generating process of the network.
For instance, Moore et al. study the addition or deletion of edges during the growth of networks with uniform or preferential attachment~\cite{moore_exact_2006}.
Considering downstream analysis, recent research has started to investigate the effects of sampling networks from random models on node centrality analysis~\cite{dasaratha_distributions_2020,avella-medina_centrality_nodate}, thus providing a possible notion of confidence of these metrics.
In contrast to our work the edge-uncertainty derives purely from the network generating process and no systematic edges errors are considered.

Note that there has also been a lot of work related to link predictability. Link prediction considers a setup in which the set of observed links in a network is used to estimate the likelihood that an unobserved link will start to exist in the future~\cite{libennowell_link-prediction_2007,clauset_hierarchical_2008,lu_toward_2015}. 
This is different to the simulation of systematic edge errors where we are not interested in predicting the temporal evolution of a network, or to reconstruct a partially observed network,
Instead we aim to quantify the implications of erroneous edges on subsequent network analysis tasks.

\subheader{Centrality in attributed and sampled networks} 
The centrality of nodes is frequently used to access the relevance and importance of entities in social networks.
For attributed social networks, the relative positioning of a group is a major determinant of how much access this group has to resources and information \cite{calvo-armengol_effects_2004,nilizadeh_twitters_2016,hannak_bias_2017}.
In this context, Karimi et al. show that a minority group in a network can become strongly (dis-)advantaged in terms of the number of network connections it can accumulate (measured in terms of node degrees), based merely on a combination of homophily and a preferential attachment mechanism \cite{karimi_homophily_2018}.

Lerman et al.~\cite{lerman_majority_2016} demonstrate that the local visibility of nodes can lead to a so called "majority illusion", a situation where individuals  systematically overestimating the prevalence of a group in the network. 
Wagner et al.~\cite{wagner_sampling_2017} observed significant differences between sampling techniques when trying capture the centrality of nodes and concluded a potential impairment of group visibility.
In contrast to our work, both studies focus on uncertainty related to having access only to parts of the (node set of the) network, rather than the impact of systematic edge errors in the network.
Other relevant studies on the effect of node-subsampling in networks include~\cite{kossinets_effects_2006,borgatti_robustness_2006,smith_online_2013}.

\subheader{Social biases, cognitive biases and systematic errors}
There are multiple biases, associated with beliefs, decision-making, memory and social desirability, which influence reporting about social connections~\cite{sen2019total,holland_structural_1973, feld_detecting_2002,brewer_forgetting_2000,marsden_network_1990,marsden_interviewer_2003}.  In the field of self-reported ego networks, research on measurement bias in respondent reports of personal networks has shown that the reported social network is usually much smaller than the actual number of relationships and that it can be influenced by the interviewer~\cite{feld_detecting_2002,van_tilburg_interviewer_1998,marsden_interviewer_2003}. A particularly timely example for a systemically biased reporting of contacts is the case of manual contact tracing in the Covid19 pandemic.  Having a high number of contacts is socially condemned, consequently individuals with a lot of interactions to others might systematically under report their number of contacts to health departments which perform contact tracing. On the contrary, individuals with a low number of contacts are more likely to report these correctly. 
For mitigation of outbreaks and effective quarantining, it is vital to capture these networks correctly which is a key challenge in current research~\cite{braithwaite_automated_2020}.

Additionally, the data collection process itself can be structurally focusing on a specific group of actors and therefore lead to a biased data availability driven by peer effects~\cite{wiese_you_2015, yang_should_2017}. 
For example, González-Bailón et al. have found that the Twitter search API over-represents the more central users and does not offer an accurate picture of peripheral activity~\cite{gonzalez-bailon_assessing_2014}. Similarly in Wikipedia clickstream data~\cite{rodi_search_2017} only transitions exceeding 10 requests are present. We would argue both of these effect are poorly modelled by random uniform missing edges.

\begin{figure*}[hbt]
    \centering
    \begin{subfigure}{0.49\linewidth}
        \begin{subfigure}{0.49\linewidth}
        \includegraphics[width=\linewidth]{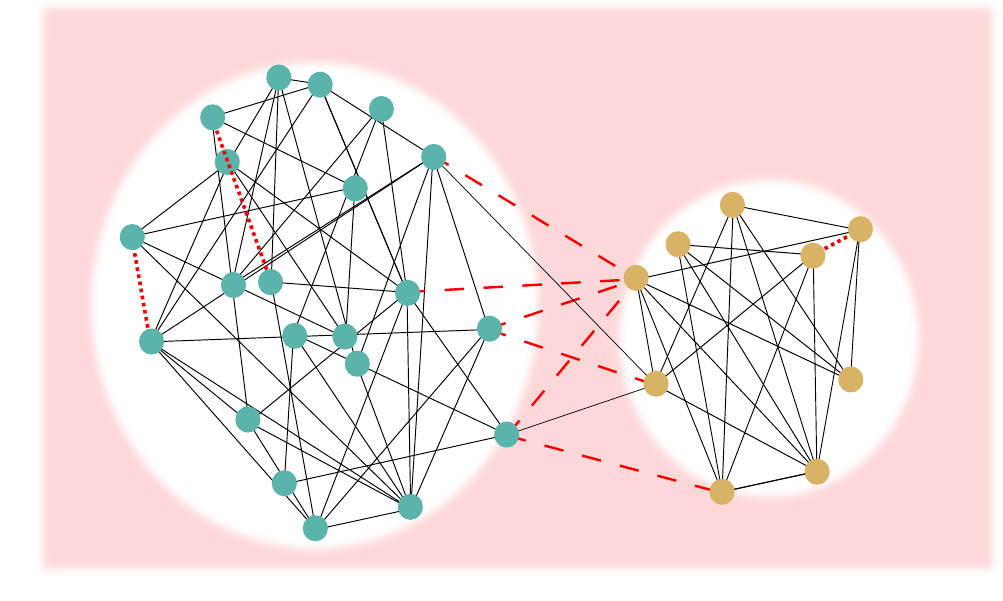}
        \caption{inter-group noise}
        \end{subfigure}
        \begin{subfigure}{0.49\linewidth}
        \includegraphics[width=\linewidth]{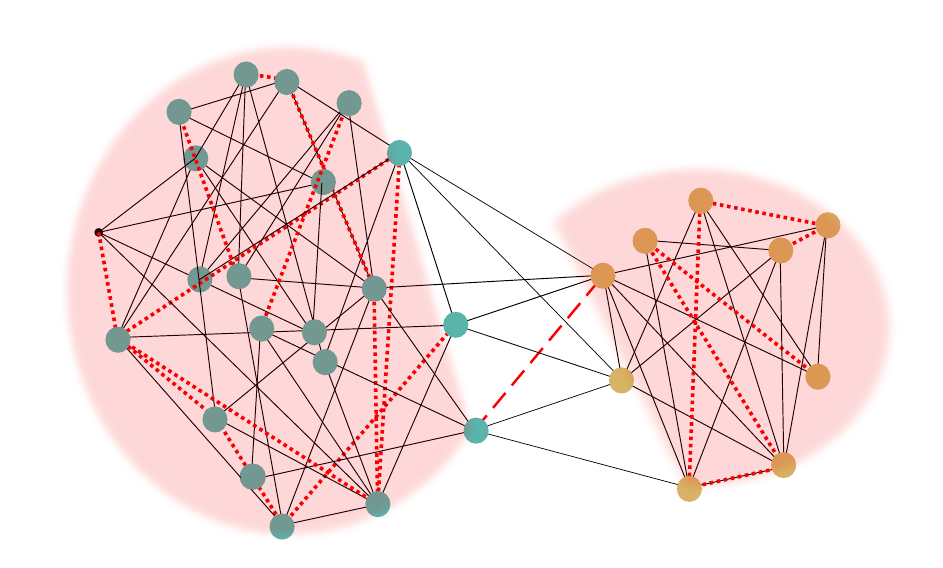}
        \caption{intra-group noise}
        \end{subfigure}
        \caption*{Group label congruence noise}
    \end{subfigure}
    \begin{subfigure}{0.49\linewidth}
        \begin{subfigure}{0.49\linewidth}
        \includegraphics[width=\linewidth]{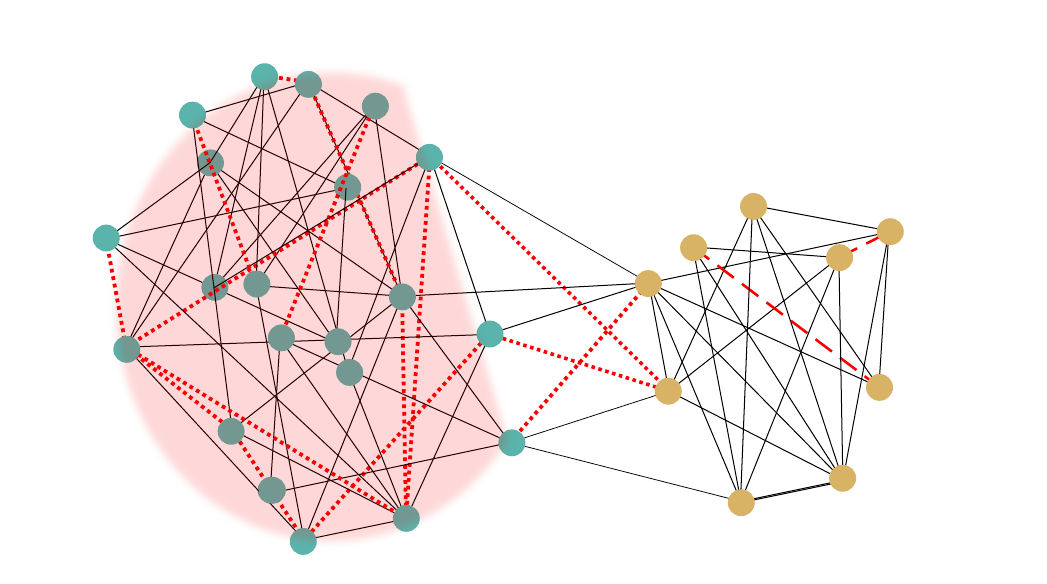}
        \caption{majority noise}
        \end{subfigure}
        \begin{subfigure}{0.49\linewidth}
        \includegraphics[width=\linewidth]{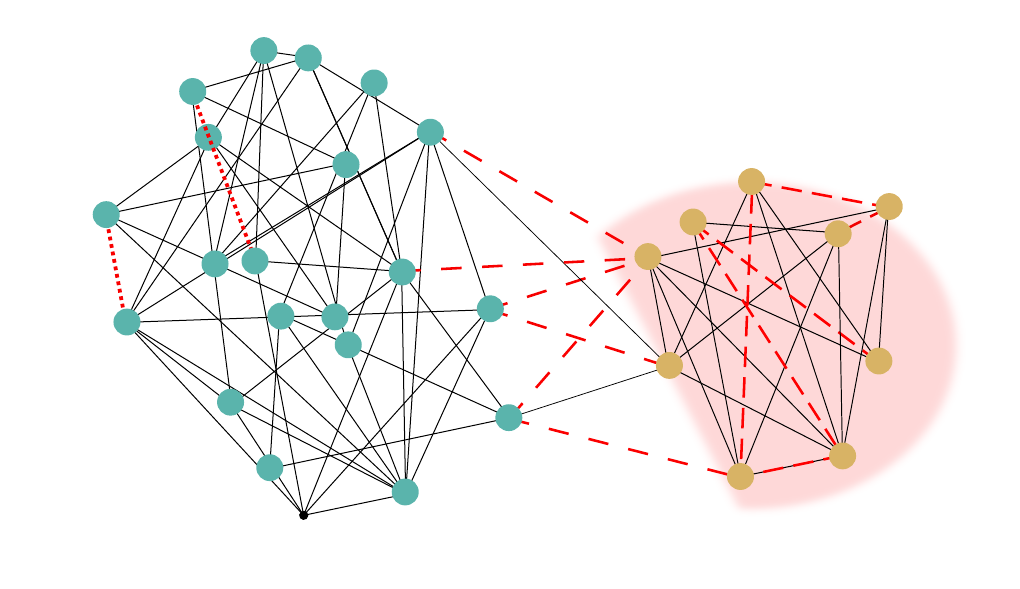}
        \caption{minority noise}
        \end{subfigure}
        \caption*{Group label specific noise}
    \end{subfigure}
    \caption{\textit{Schematic representation of node attribute-based noise.}
      We visualize the four different cases for node attribute-based noise for a homophilic network.  In Figure (a, b) we can see the two cases of group label congruence noise, inter-group noise on the left and intra-group noise on the right.  In Figure (c, d) we illustrate group label specific noise. If we define the majority as the specific group of interest, we can see majority affiliated noise on the left and minority affiliated noise on the right. Note that here we only show homophilic networks while we also consider neutral/heterophilic networks in our experiments.}
    \label{fig:metadatanoise}
\end{figure*}
\section{Systematic error model}
\label{sec:systematic_error_model}

The main contribution of our work is the introduction of a model for simulating systematic edge noise, with which we explore different (hypothetical) biases in data collection and interpretation. 
For instance, due to a social desirability bias one social group might over- report or under report connections with another group. Alternatively, due to cognitive availability biases, people may remember interactions with closely related others more \cite{bell_partner_2007,smieszek_collecting_2012,calloway_accuracy_1993}. They might also tend to report more connections to nodes with a high visibility in their social network.
We model such social, cognitive and technical biases by systematically adding or removing edges to a network. 
While our model equally allows for the erroneous addition or removal of edges (e.g., over- and under reporting), in our numerical experiments we concentrate on edge omissions, i.e., edges that are not present in the observed network, but are actually present in the latent social network.
For the mathematical details of our model we refer to Section \ref{sec:methods}.

In the following, we describe the different ways in which we model systematic noise.
We distinguish between two types of critical information for systematic edge errors: (i) node attribute data, which is extrinsic to the network, and (ii) local network structure properties, which is information that is intrinsic to the network. 
We encode node attributes in the form of continuous or discrete node labels $l_i$.
For instance, the label $l_i$ may correspond to the height of a person or its membership in a particular social group.
We will focus on categorical node attributes representing two groups in the following discussion, although our models can be applied to other discrete or continuous node attribute data as well (cf. Section~\ref{ssec:additional_considerations} for additional discussion).

Note that the two different types of bias in our model differ in the type of information they are based on, but not necessarily in their effect on the network data.
This can imply that a node attribute-based bias and a structure-based bias have the same effect, even if two different biases are modelled. 
Such an effect arises, e.g., if the node labels are aligned with the respective structural properties. 

\subsection{Baseline model}
Removing edges uniformly at random has previously been studied \cite{borgatti_robustness_2006,kossinets_effects_2006} and serves as our baseline model.
In the baseline model, errors are independent and identical for each edge.
This kind of measurement error may be relevant in settings in which all observed noise is technical, e.g., if we measure social contacts via some proximity sensors, which may independently fail to detect a contact with a certain probability ~\cite{dubois_effect_2012,martin_random_2006}.
Moreover, the baseline model enables us to better gauge and compare the effects of systematic errors.

\subsection{Attribute-based noise}
In the context of social networks, systematic errors are often driven by node attributes, e.g. membership in certain groups (based on ethnicity, gender, etc.).
For a network split into multiple groups, we may for instance observe certain forms of in-group/out-group bias or social desirability bias, where a connection to a particular group may be less or more often reported (e.g. in-group favoritism \cite{everett_preferences_2015}). 
Here we are particularly  interested in two scenarios, which we call \emph{group label congruence} and \emph{group label specific} noise.

\paragraph{Group label congruence noise (intra / inter group)}
This setting describes classical inter-group (between groups) vs. intra-group (within group) effects. The characteristic that determines the noise level is whether the nodes incident to an edge have the same or distinct labels, i.e., belong to the same group or not.

\paragraph{Group label specific noise (majority/minority)}
This setting models a scenario where edges attached to nodes of a particular group, are significantly more or less affected by noise. 
For instance, edges connecting to a minority group may be under-reported.
In our model, the noise level thus depends on whether one of the nodes on the end of an edge is part of a specific group.

Figure \ref{fig:metadatanoise} provides illustrations for group label congruence and group label specific edge noise, for the common scenario in which the network consist of two groups: a majority and a minority.
This setup will be the main focus for our numerical experiments.

\subsection{Network structure-based noise}
Certain systematic errors are intrinsically coupled to the structure of the latent network itself. For instance, humans tend to recall information better that is more readily available in their memory, e.g., because the information is more recent or associated to an event of higher (perceived) importance \cite{bell_partner_2007,smieszek_collecting_2012,calloway_accuracy_1993}. We propose two forms of systematic noise (Jaccard and Centrality noise) which could be used to model these effects as a function of network topology.

\paragraph{Jaccard noise (availability bias)}
When working with data from self-reported, ego centered networks, frequent interactions will be recalled more likely than rare interactions. If two nodes share a lot of neighbours in a social network, i.e., their social circles overlap significantly, this can indicate more frequent interactions between the two nodes.
Following the logic of an availability bias, in our model edges connecting two nodes with a high neighbourhood overlap are thus less likely to be forgotten. 
Vice versa, edges between nodes whose neighbourhoods have little overlap may be forgotten more easily. 

\paragraph{Centrality noise (desirability bias)}
Alternatively, interactions with highly central nodes in the network might be perceived as more important, and interactions to these individuals would accordingly be remembered better when reporting ties.
Social desirability may amplify this even further, leading to a higher probability of observing interactions with central nodes at the core of the network.
When measuring network data, the core of the network might thus be captured more accurately than the periphery.
We thus model centrality noise by retaining more edges between nodes with high degree and dropping edges that connect low degree nodes with higher probability.

\section{Mathematical error modeling}
\label{sec:methods}
This section outlines the mathematical details of our edge error model.
In this work we consider networks represented by simple undirected graphs $\mathcal{G}(\mathcal V, \mathcal E)$ with a vertex set $\mathcal V = \{1,\dots,N\}$ and an edge-set $\mathcal E=\{\{i,j\}:~i,j\in~\mathcal V\}$. 
We can represent such a graph algebraically by an adjacency matrix $A \in  \mathbb{R}^{N \times N}$ with entries $A_{ij} = 1$ if $\{i,j\} \in \mathcal E$ and $A_{ij} = 0$ else.
As the graph is undirected $A$ is a symmetric matrix. 

When observing a network, we will generally obtain an (observed) adjacency matrix $A^o$, which is usually an error prone observation of an underlying latent network $A$.
If we observe an edge in $A^o$ we may have (i) a true positive observation, meaning that the edge was indeed present in the latent network or (ii) a false positive observation, where we observe an edge, when there should have been none.
We model these cases by defining the probability of an observed true positive edge $A_{ij}^o$ as $P_{ij} = \mathbb{P}(A_{ij}^o=1|A_{ij}=1)$.
The true positive probability $P_{ij}$ can be interpreted as the probability for a latent edge to be retained in the observed network, and we thus alternatively refer to it as retain probability.
Analogously, if we do not observe an edge, we may have (i) a true negative observation, where we observe no edge and there was indeed none, or (ii) a false negative observation, i.e., we may not observe an edge when there is one in the latent network.

For edge errors described by statistically independent $\{0,1\}$ Bernoulli random variables, we can encode the marginal probabilities of the observed true positives in a symmetric $n\times n$ matrix $P$. 
We denote a sample from this distribution over matrices by $S_P$. The adjacency matrix of the observed network may then be expressed as compactly as:
\begin{equation}
    A^o = A \odot S_P, 
\end{equation}
where $\odot$ denotes the element-wise (Hadamard) product. 
To put it in other words, each edge $(i,j)$ present in the latent network is retained with probability $P_{ij}$. This means that once the probabilities are fixed the edges are dropped independently. 

In this work we focus on the particular case of systematic missing edges modelled by being drawn independently. Generalizations such as edge addition or edges being dropped in a correlated way are also possible with minor adaptation.
For instance, we can describe the observed adjacency matrix of a model with independent edge additions analogously to above as:
\begin{equation}
    A^o = A \odot S_P + A^c \odot S_Q,
\end{equation}
where $A^c$ denotes the adjacency matrix of the complement graph and $S_Q$ a $\{0,1\}$ sample drawn according the matrix of false positives.
%
In the following subsections we discuss in detail how (systematic) edge-errors can shape the error probability matrix $P$.

\subsection{Baseline model}

In the baseline model, every edge that is present in the latent network is retained with constant probability $p$.  
This leads to a (constant) noise probability matrix:
\begin{equation}
     P_{ij} = p
\end{equation}
Note that low values of $p$ correspond to a large impact on the observed network.

\subsection{Node attribute-based noise}\label{ssec:attribute_noise}

Given that each node $i$ in the network has an associated categorical label $l_i = 1,\ldots,\ell$,
we now assume that noise is solely a function of the pair of labels involved in each edge. 
We thus write
\begin{equation}
    P_{ij} = \Omega_{l_i l_j}
\end{equation}
with $\Omega_{mn}$ encoding the probability that an edge between nodes with label $m$ and $n$ is retained. 
%
In the following we investigate how the matrix $\Omega$ can be chosen to account for different kinds of noise.

\paragraph{Group label congruence noise (intra / inter group)}
For group label congruence noise we assume that it is mostly important whether nodes have the same label (intra) or not (inter).
In this case, the matrix $\Omega$ will be structured as:
\begin{equation}
\Omega_{m n} = \begin{cases}p_\text{intra} & m=n,\\
  p_\text{inter} & \text{otherwise}. \end{cases}
\end{equation}
where $p_\text{intra}$ describes the probability of retaining an edge if two nodes are in the same group (have the same node label), and $p_\text{inter}$ describes the probability of retaining an edge if two nodes are in different groups.
We can distinguish between $p_\text{intra} < p_\text{inter}$ in which intra-group connections are less likely to be retained, and $p_\text{intra} > p_\text{inter}$ in which the same is true for inter-group edges.

\paragraph{Group label specific noise (majority/minority)}
For label specific noise we assume that edges which are attached on either end to one particular label $l$ are more strongly affected by noise.
The matrix $\Omega$ will be structured as:
\begin{equation}
\Omega_{m n} = \begin{cases}p_l & m=l \lor n=l,\\
  p_\text{standard} & \text{otherwise}. \end{cases}
\end{equation}
This means, edges that are on either end connected to nodes of that particular class $l$ are retained with probability $p_l$ while all other edges are retained with a probability $p_\text{standard}$.
\subsubsection{Parameter specification of $\Omega$ in numerical experiments}
In our simulations in Section \ref{sec:experimental _results}, we explore four different node attribute-based noise types (i) inter-group noise, (ii) intra-group noise, (iii) minority specific noise, and (iv) majority specific noise. The exact parametrization of the introduced label-based error matrix $\Omega$ which captures these four settings can be found in Table \ref{tab:rho}.

\begin{table}[b!]
\caption{\textit{Specific parametrization of label-based error matrix $\Omega$ used for numerical experiments.} { We model four different noise types. We use the retain parameter $\rho$ to control the noise probability of edges. For inter-group noise, e.g., we retain an edge that connects the two classes with probability $\rho$, while we retain edges within a class with 90\% probability.}}
\centering
\begin{tabular}[b]{ l | c | c | c |c }
noise type & intra & inter& majority & minority\\  \hline
$\Omega$ &
$
\begin{pmatrix}
\rho & 0.9\\
0.9 & \rho
\end{pmatrix}
$
&
$\begin{pmatrix}
0.9 & \rho\\
\rho & 0.9
\end{pmatrix}$
&
$\begin{pmatrix}
\rho & \rho\\
\rho & 0.9
\end{pmatrix}$
&
$\begin{pmatrix}
0.9 & \rho\\
\rho & \rho
\end{pmatrix}$
\end{tabular}
\label{tab:rho}
\end{table}

\subsection{Network structure-based noise}\label{ssec:positiion_noise}

To model the aforementioned biases and systematic errors, we consider the structured error matrix $P$, such that the probability $P_{ij}$ of an edge $\{i,j\}$ to be retained is a function $f(\{i,j\},\mathcal G)$ of the position of the specific edge in the network.

We argue that local properties are of particular interest for modelling social and cognitive biases as most global information is typically not available to individuals. 
Consequently, the specific functions $f(\cdot)$ considered below depend only on information that is locally available to the two nodes adjacent to the edge (or can be computed based on local information).
To investigate the effects of such kind of noise, we consider two specific scenarios in the following.

\paragraph{Jaccard noise (availability bias)} 
As one of the concrete examples of structure-based noise we consider a model that can capture the effects of availability bias.
Specifically, the probability of an edge ${i,j}$ to be retained or falsely included is a function of the overlap of the neighborhoods between the two endpoints $i,j$. 
In line with the idea of availability bias, if the two endpoints have a very similar neighborhood, then the probability of retention is high.
We model the availability bias by retaining an edge with a probability that is given by the Jaccard similarity of the (augmented) neighborhoods of the two endpoints of the edge:
\begin{equation}
    P_{ij} = \left( \frac{|\mathcal{N}_i \cap \mathcal{N}_j|}{|\mathcal{N}_i \cup \mathcal{N}_j|} \right)^\alpha =: \mathcal J^\alpha
    \label{eq:jaccard}
\end{equation}
To avoid removing all edges in the case of almost bipartite networks, here we consider each node $i$ to be part of its own neighborhood as well, i.e., $\mathcal{N}_i =   \{i\}\cup\{k : \{k,i\} \in \mathcal E \}$. $\alpha$ is a scaling parameter that we can choose to modulate how aggressive edges will be dropped. For $\alpha =0$ all edges will be retained, for larger $\alpha$ we more aggressively remove edges whose endpoints have a low product of degrees (relative to other node pairs in the graph).
We call the noise of type Eq.~\eqref{eq:jaccard} \emph{Jaccard noise}.
In addition we define the \emph{inverse Jaccard noise} via $P_{ij} =\left( 1 - \mathcal J \right)^\alpha$, which models a situation in  which an edge between two similar nodes is more likely to be dropped. 

\paragraph{Centrality-based noise (social desirability)}
Our framework also enables us to model a form of availability 
bias that is driven by the (perceived) importance of the nodes, e.g., by making the probability of a true positive proportional to the centrality of the two endpoints ($P_{ij} \propto c(i)c(j)$ where $c(i)$ is the centrality of node $i$).

The specific scenario we consider in the following is again based on local information.
We model a setting in which the probability for edges to be omitted will be larger in the periphery than in the core of the network, where we define core/periphery based on the degrees of the nodes.
Specifically let $d_i =\sum_j A_{ij}$ denote the degree of node $i$.
We then describe our matrix of retain probabilities via:
\begin{equation}
    P_{ij} = \left(\frac{d_{ij} - \min_{ij \in  \mathcal E} d_{ij}}{\max_{ij \in  \mathcal E} d_{ij} - \min_{ij \in  \mathcal E} {d_{ij}}}\right)^\alpha =: \mathcal D_{ij}^\alpha,
    \label{eq:structure}
\end{equation}
where $d_{ij} = \log(d_i) + \log(d_j)$, and $\alpha$ has the same function as for Jaccard noise.
We call~\eqref{eq:structure} \emph{centrality-based noise}.
As before, we define the \emph{inverse centrality-based noise} via $P_{ij} = \left(1- \mathcal D_{ij} \right)^\alpha$.

\section{Experimental Results}
\label{sec:experimental _results}

In our experiments, we focus on attributed networks with binary labels which represent two groups with unequal group sizes. 
We investigate how the systematic errors we introduced in section~\ref{sec:systematic_error_model} affect the representation of a minority in degree-based node rankings in networks with different degree of homophily. 
Our overall goal here is to elucidate possible effects of interactions of specific systematic bias with the network topology.  
Specifically, we consider the fraction of minority nodes in the top $k$ nodes of the ranking.
We perform experiments on both synthetic and real-world networks. 

\subsection{Experiments on synthetic networks}
\label{ssec:experiments_synthetic}

\subsubsection{Experimental setup}
We first investigate the effects of the different noise types on synthetic networks in which we can control the homophily of the two groups. 
We use an adaptation of the Barabasi-Albert (BA) model, introduced in~\cite{karimi_homophily_2018}. 
%
%
The model uses a preferential attachment mechanism where the attachment probability for a new node of label $a$ to attach with a node of label $b$ is additionally skewed by their homophily $h_{ab}$.
Note that we only consider the case of symmetric homophily for binary labels, which means that $h_{00}=h_{11}=h$ and $h_{01}=h_{10}=1-h$.
Now $h=0$ indicates an entirely heterophilic network i.e. nodes of group 0 only connect to nodes of group 1 and vice versa.  The case of $h=0.5$ is equivalent to the standard BA model where node labels play no role in the formation of the network and $h=1.0$ is a completely homophilic network where nodes of type $l$ only connect to nodes with similar label $l$.

In our numerical experiments, we focus on binary minority / majority labels with a fixed minority size of $10\%$ of the nodes.
In Section~\ref{sensitivity} in the appendix we assess the robustness of our results with respect to this choice.
We generate networks with $N=10000$ nodes by adding $m=5$ edges in the modified preferential attachment scheme~\cite{karimi_homophily_2018}.
The different noise types are then applied to each network. 
We include both simulations of independent uniform noise and calculations without edge-errors as baselines for the systematic error scenarios.

We explore four different node attribute-based noise types (i) inter-group noise, (ii) intra-group noise, (iii) minority specific noise, and (iv) majority specific noise. We use a retain parameter $\rho$ to control the noise probability of edges. For inter-group noise, we retain an edge that connects the two classes with probability $\rho$, while we retain edges within a class with 90\% probability, vice versa for intra-group noise. For minority/majority specific noise, the retain probability of an edge connecting with at least one node of the specific group given by $\rho$, and only if both nodes belong to another group, the edge is retained with a probability of $90\%$. Mathematically, this is modeled through different structures of a label-based error matrix (see Section~\ref{ssec:attribute_noise} for further details). 

Additionally we investigate network structure-based noise types: Jaccard and inverse Jaccard noise, as well as centrality-based and inverse centrality-based noise. 
The strength of the noise is controlled by an exponent parameter $\alpha$ (see Eqs.~\ref{eq:jaccard},\ref{eq:structure} in Section~\ref{ssec:positiion_noise}). For low values of $\alpha$ edges are retained more likely while for large $\alpha$ fewer and fewer edges are retained. This in contrast with $\rho$ where low values correspond to strong noise.
For each parameter setting, we generate 10 networks.

\begin{figure*}[ht!]
	\begin{center}
	\begin{subfigure}[c]{2.0\columnwidth}
			\includegraphics[width=\textwidth]{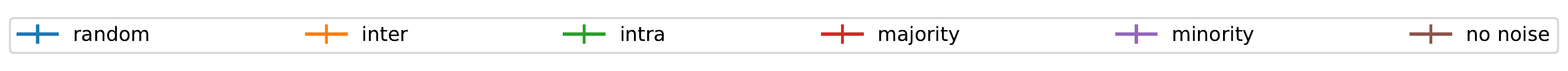}
		\end{subfigure}
		\begin{subfigure}[b]{0.452612\columnwidth}
			\includegraphics[width=\textwidth]{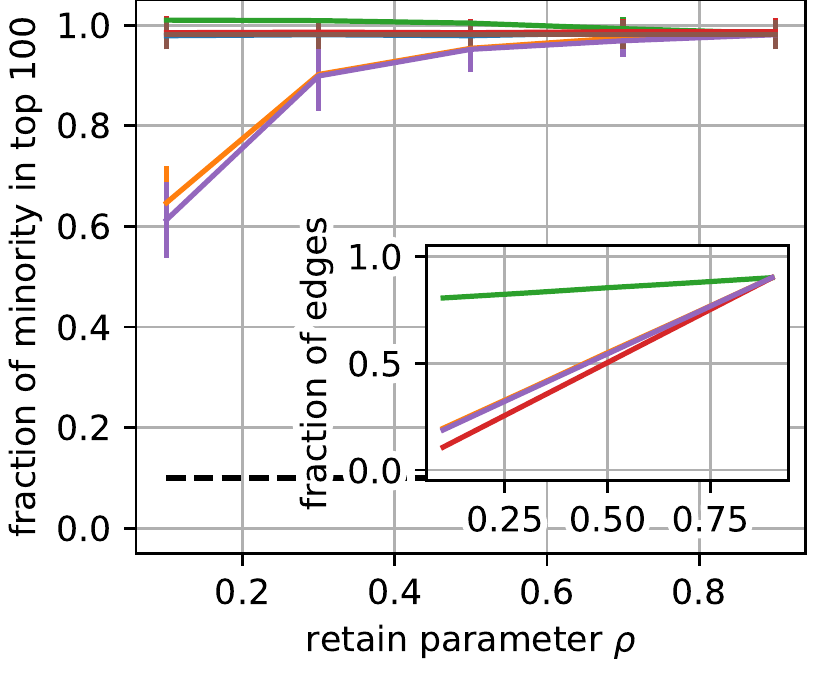}
			\caption{h=0.10 (heterophilic)}
		\end{subfigure}
		\begin{subfigure}[b]{0.386847\columnwidth}
			\includegraphics[width=\textwidth]{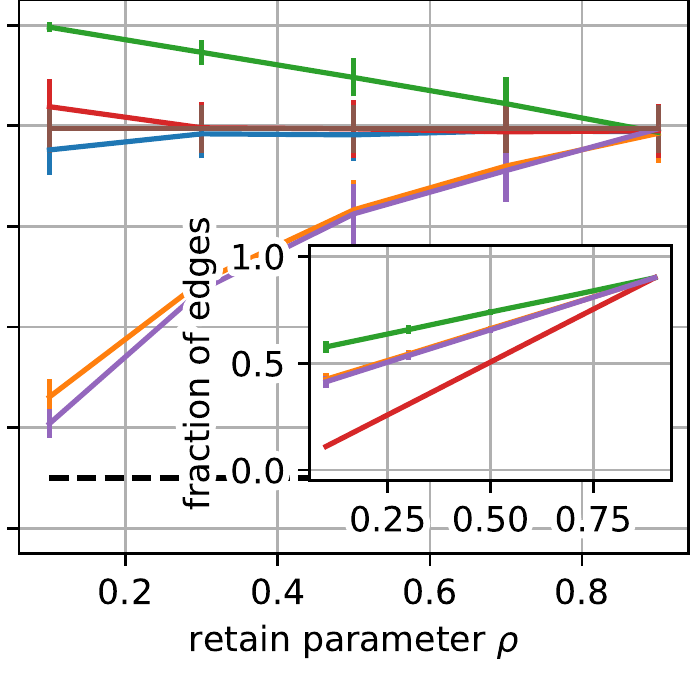}
			\caption{h=0.25}
		\end{subfigure}
		\begin{subfigure}[b]{0.386847\columnwidth}
			\includegraphics[width=\textwidth]{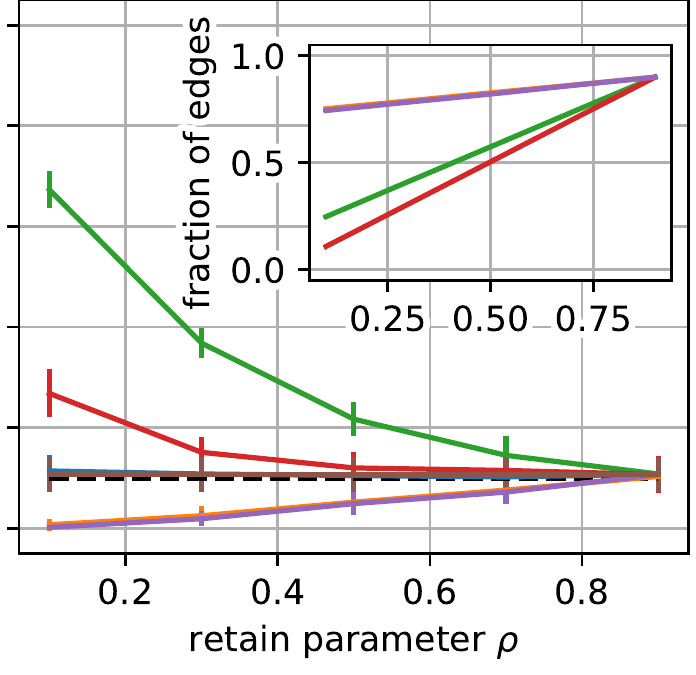}
			\caption{h=0.50 (neutral)}
		\end{subfigure}
		\begin{subfigure}[b]{0.386847\columnwidth}
			\includegraphics[width=\textwidth]{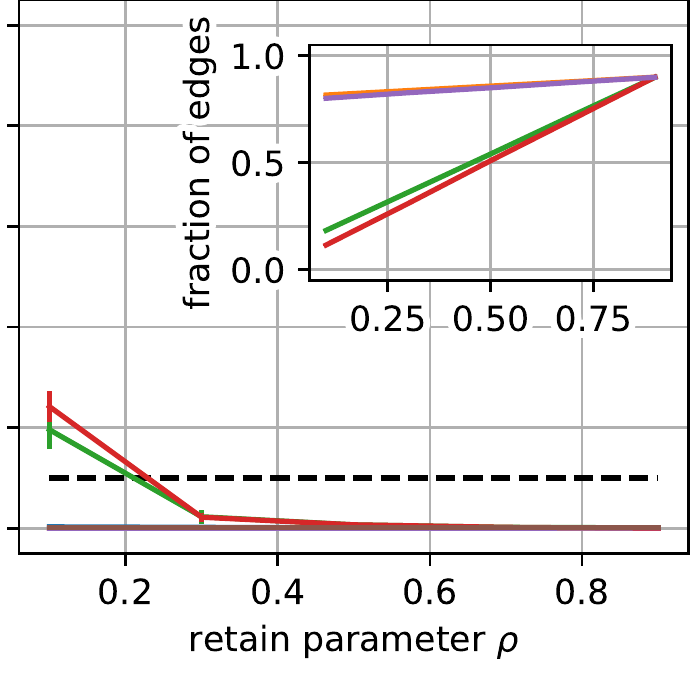}
			\caption{h=0.75}
		\end{subfigure}
		\begin{subfigure}[b]{0.386847\columnwidth}
			\includegraphics[width=\textwidth]{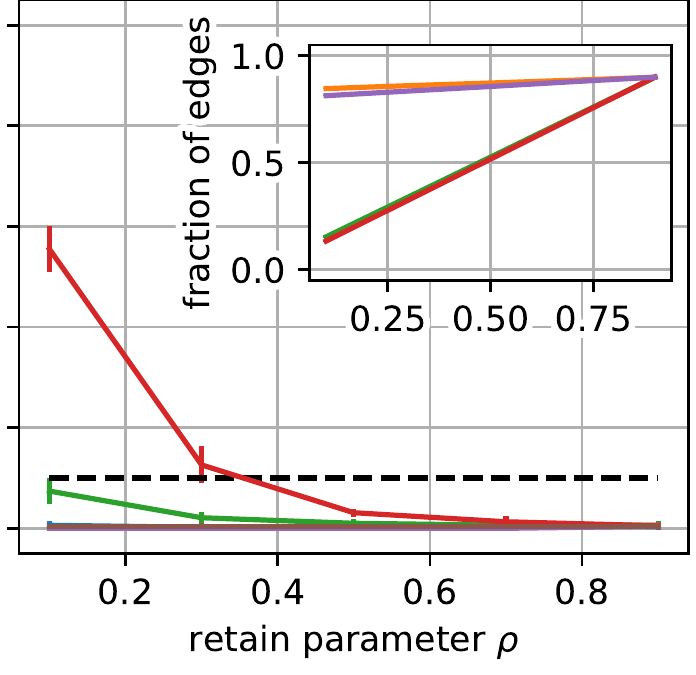}
			\caption{h=0.90 (homophilic)}
		\end{subfigure}
	\end{center}
\caption{\textit{Impact of node attribute-based noise on the representation of minority nodes in rankings on synthetic networks.}
{We visualize the fraction of minority in the top 1\% (top 100) nodes as a function of the retain parameter $\rho$. 
Note that low values of $\rho$ correspond to a strong noise (noise decreases from left to right in every subplot).
We  increase the homophily parameter $h$ from the leftmost to the rightmost panel. The main plots show the representation of the minority in a degree-based ranking while the insets show the impact of $\rho$ on the number of edges in the network. 
We can see that in heterophilic regimes the minority representation is already affected by relatively weak noise. 
In contrast, the representation in homophilic regimes is only affected for much stronger noise.
This shows that node attribute-based, systematic noise can have a variety of effects which depends on the homophily of the network.}
}
\label{fig:retain}
\end{figure*}
\begin{figure*}[ht]
	\centering
	\begin{subfigure}[c]{2.0\columnwidth}
			\includegraphics[width=\textwidth]{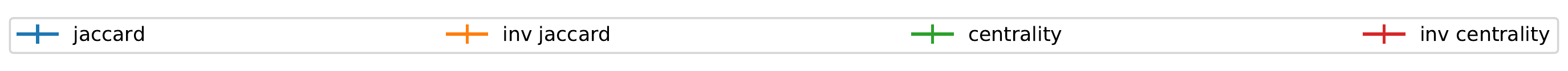}
		\end{subfigure}
		\begin{subfigure}[b]{0.452612\columnwidth}
			\includegraphics[width=\textwidth]{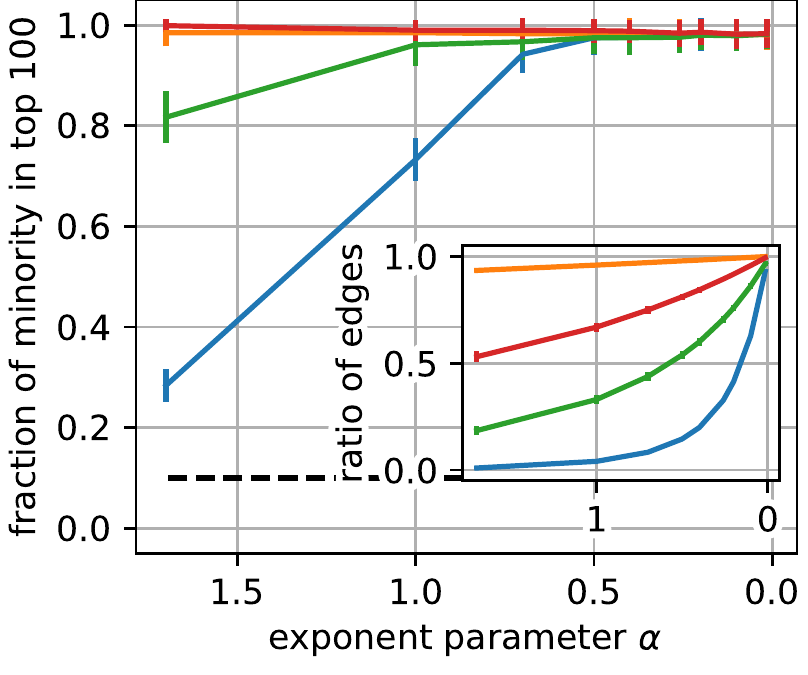}
			\caption{h=0.10 (heterophilic)}
		\end{subfigure}
		\begin{subfigure}[b]{0.386847\columnwidth}
			\includegraphics[width=\textwidth]{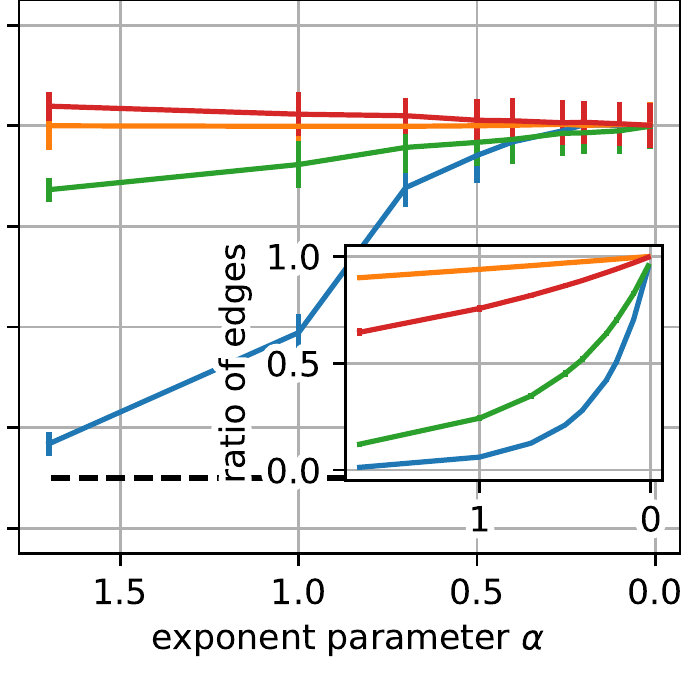}
			\caption{h=0.25}
		\end{subfigure}
		\begin{subfigure}[b]{0.386847\columnwidth}
			\includegraphics[width=\textwidth]{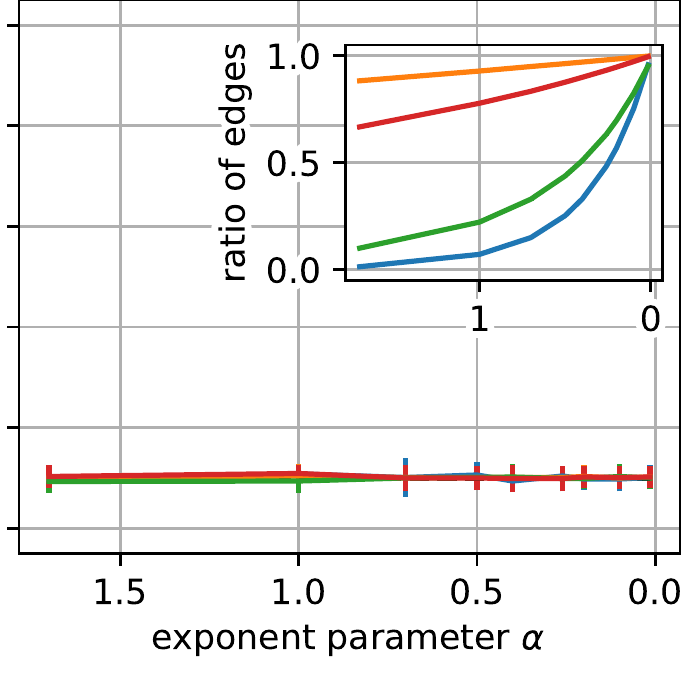}
			\caption{h=0.50 \smaller{(neutral)}}
		\end{subfigure}
		\begin{subfigure}[b]{0.386847\columnwidth}
			\includegraphics[width=\textwidth]{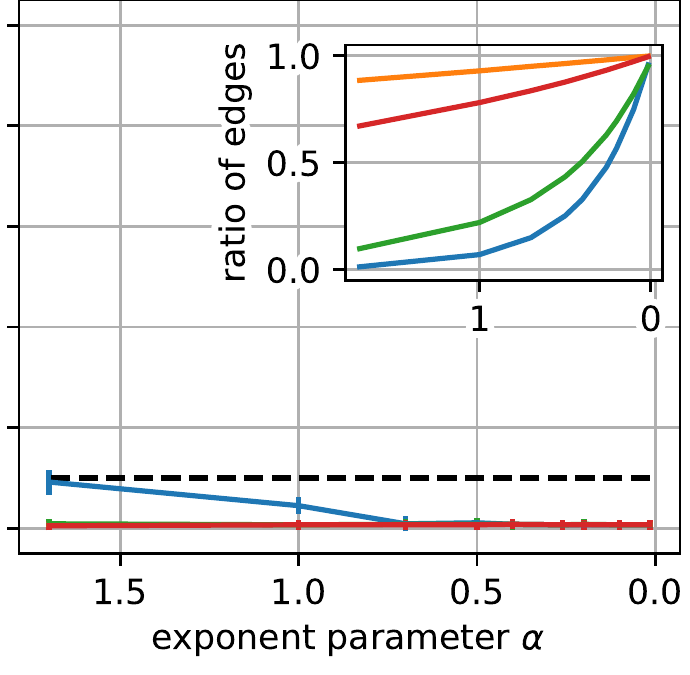}
			\caption{h=0.75}
		\end{subfigure}
		\begin{subfigure}[b]{0.386847\columnwidth}
			\includegraphics[width=\textwidth]{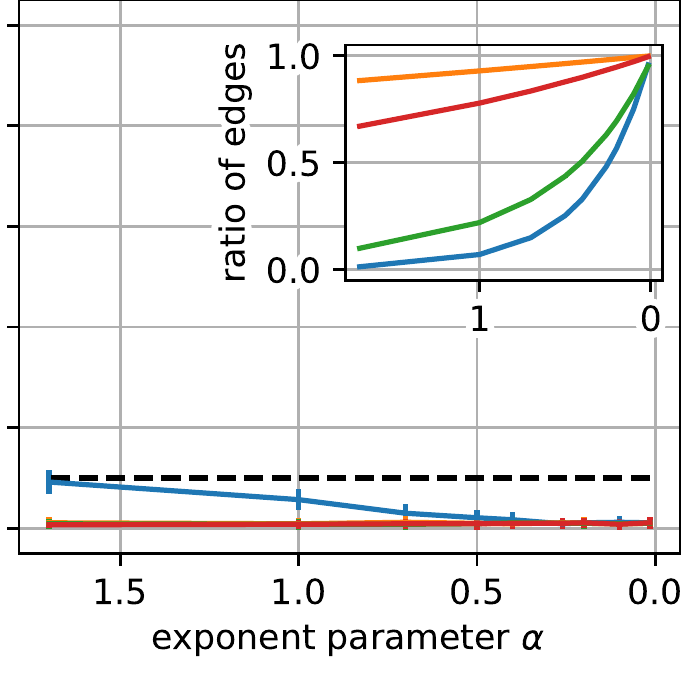}
			\caption{h=0.9 (homophilic)}
		\end{subfigure}
\caption{\textit{Impact of structure-based noise on the representation of minority nodes in rankings on synthetic networks.} {Similar to Fig.~\ref{fig:retain} we visualize the fraction of the minority as a function of the noise strength represented via $\alpha$. Larger values of $\alpha$ correspond to higher noise-values (noise decreases from left to right in every subplot).
The insets show the impact of noise on the number of edges in the network. 
As the noise is not aligned with the relative group connectivity regulated by the homophily parameter $h$, we can see that the general amount of edges dropped for each noise type is relatively independent of $h$.  Although centrality-based noise leads to omitting comparatively many edges, the impact of centrality noise on the minority in the degree ranking is not as strong as Jaccard noise. }}

\label{fig:retain2}
\end{figure*}

\subsection{Results for synthetic networks}
In Figure~\ref{fig:retain} and \ref{fig:retain2} we display the simulation results of the effects of attribute-based and structure-based noise. In all simulations, we already observe interesting effects for the noise-free setting, which are caused by the homophily of the network independent of noise and have been discussed by Karimi et al~\cite{karimi_homophily_2018}: In a homophilic regime (low $h$, see Fig.\ref{fig:retain} and \ref{fig:retain2} d,~e), we find a general under-representation of the minority.
This effect is due to the relative size of the majority which prefers interactions among itself. In contrast, in a heterophilic regime ($h\le0.25$, see Fig.~\ref{fig:retain} and \ref{fig:retain2} a,~b), the number of inter-group edges is high, resulting in a strong over-representation of the minority in the degree ranking in the noise free case, relative to the actual group-sizes~\cite{karimi_homophily_2018}. Throughout our analysis we interpret this over/under representation as the baseline which is indicated through the "no-noise" line.

\subsubsection{Node attribute-based noise}
The results for the effects of attribute-based noise in Figure~\ref{fig:retain} show contrasting effects for homophilic and heterophilic networks for different types of noise. 
We find that in the case of a heterophilic network (Fig. \ref{fig:retain} a,b), systematic noise can have large and opposing effects compared to the error-free and random baselines.
Noise types that remove preferably inter-group edges reduce the representational advantage of the minority. 
Vice versa, dropping intra-class edges further increases the over-representation of the minority, as the significance of the minority in the degree-based ranking is not due to intra-class edges, but due to a large number of inter-class edges.

In contrast, we find that the under-representation of the minority in homophilic networks is stable with regard to different noise types and over a broad range of retain probabilities (Fig. \ref{fig:retain} d,e).
The minority can only gain some significance in the ranking for very low edge retain probabilities in the majority noise and intra-group noise scenarios.
In these two scenarios the majority will proportionally loose many more edges than the minority.

If the minority and majority nodes are distributed randomly in the network (corresponding to no homophily, $h=0.5$) (see Figure ~\ref{fig:retain} c), the  relative representation of the minority/majority in the noise-free degree-based ranking is only dependent on the relative group sizes ($10\%$ minority).
Accordingly, we find opposing effects of different noise types with respect to this baseline for $h=0.5$: intra-group and majority noise lead to an over representation of the minority, inter-group noise and minority noise lead to an under representation.

\subsubsection{Network structure-based noise}
In our second set of experiments, shown in Fig.\ref{fig:retain2}, we focus on network structure-based noise.

For the extended BA model, Jaccard noise mostly penalizes edges that connect low degree with high degree nodes.
In a heterophilic setup (Fig. \ref{fig:retain2} a,b), most of these edges are inter-group edges and Jaccard noise thus behaves similarly to inter-group noise in the case of attribute-based noise.  
In contrast, in homophilic regimes (Fig. \ref{fig:retain2} d,e), most of the edges that connect low degree with high degree nodes are intra-group edges and the Jaccard-noise is thus similar to the case of intra-group noise.
Depending on the homophily parameter, we therefore either penalise the minority or majority, as this type of noise is not dependant on the labels as in the case of attribute-based noise, but derived from the network structure itself. 

Inverse Jaccard noise mostly drops edges within strong communities i.e. within densely connected subsets of nodes which are not present in the extended BA model. 
This leads to an omission of very few edges and almost no shift in representation.

This highlights an important difference of our two types of noise: attribute-based noise is derived from groups which are based on the meta data of the nodes, whereas the structure-based noise (Jaccard noise in particular) is dependant on the community structure of the network. As community structure of the network and the metadata of different groups do not necessarily have to align \cite{peel_ground_2017}, this can lead to different effects depending on how the structure and meta-data correspond in the particular network of interest.

For centrality noise, the effects we observe are not as strong as for Jaccard noise. Centrality noise mostly drops edges connecting low to low/mid degree nodes. 
This leaves the high degree nodes represented in rankings primarily untouched. 
Thus, dropping comparatively many edges has no effect in homophilic regimes where the majority is in advantage. We remark that the above finding does not imply that centrality-based noise has no effect on the network. 
Rather, for centrality-based noise the periphery-periphery edges are mostly affected, and these do not show up in the ranking measure we consider here.
In heterophilic regimes a substantial amount of the in-edges of top nodes in either group is supplied by nodes of the other group. 
Since minority nodes have a higher \emph{average} degree this implies that links from the minority to high degree nodes in the majority are more likely being retained. This leads to a reduced over-representation of minority nodes. 
This effect is not reversed for inverse centrality noise.
Due to the specific structure of the BA model there are far less edges that connect two very high-degree nodes and thus very few edges have a high probability of getting omitted (cf. the insets of Fig.\ref{fig:retain2}). 
Thus, the inverse centrality noise leaves the network mostly untouched.

We can conclude from our analysis on synthetic networks that systematic biases can have very different effects on the ranking outcomes depending on the joint effect of degree of homophily and type of systematic error. We find that in heterophilic networks, majority/minority representations in rankings are very sensitive to the type of edge error present. 
In contrast, in homophilic networks we find that minorities are at a disadvantage regardless of the type of error present.

\begin{figure*}[hbt]
	\begin{center}
	\begin{subfigure}[c]{2.0\columnwidth}
				\includegraphics[width=\textwidth]{figures/legend10.pdf}
		\end{subfigure}
		\begin{subfigure}[b]{0.452612\columnwidth}
			\includegraphics[width=\textwidth]{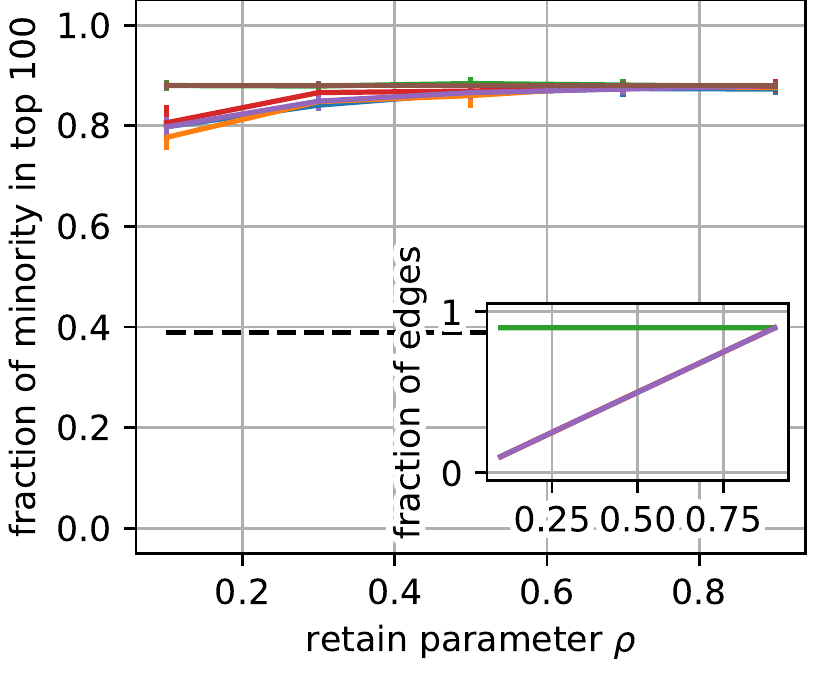}
			\caption{brazil \tiny{(strongly heterophilic)}}
		\end{subfigure}
		\begin{subfigure}[b]{0.386847\columnwidth}
			\includegraphics[width=\textwidth]{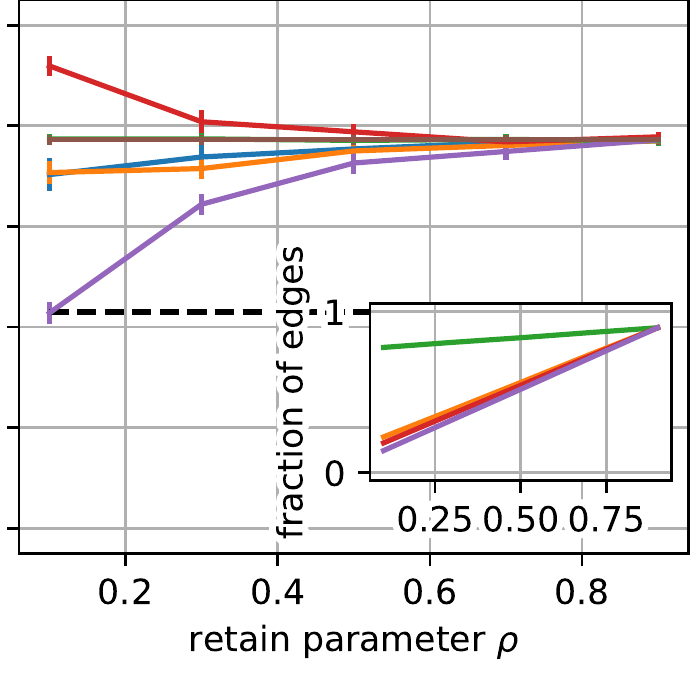}
			\caption{pok \tiny{(strongly heterophilic)}}
		\end{subfigure}
		\begin{subfigure}[b]{0.386847\columnwidth}
			\includegraphics[width=\textwidth]{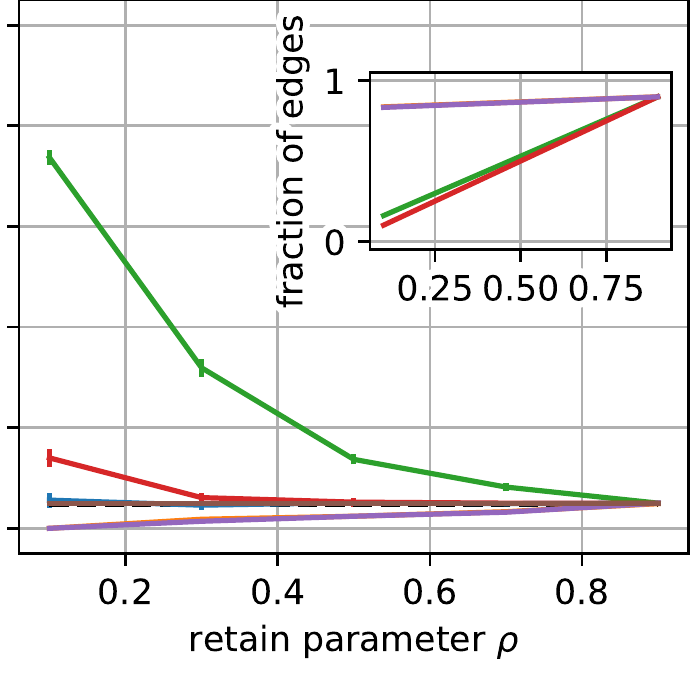}
			\caption{github \tiny{(neutral)}}
		\end{subfigure}
		\begin{subfigure}[b]{0.386847\columnwidth}
			\includegraphics[width=\textwidth]{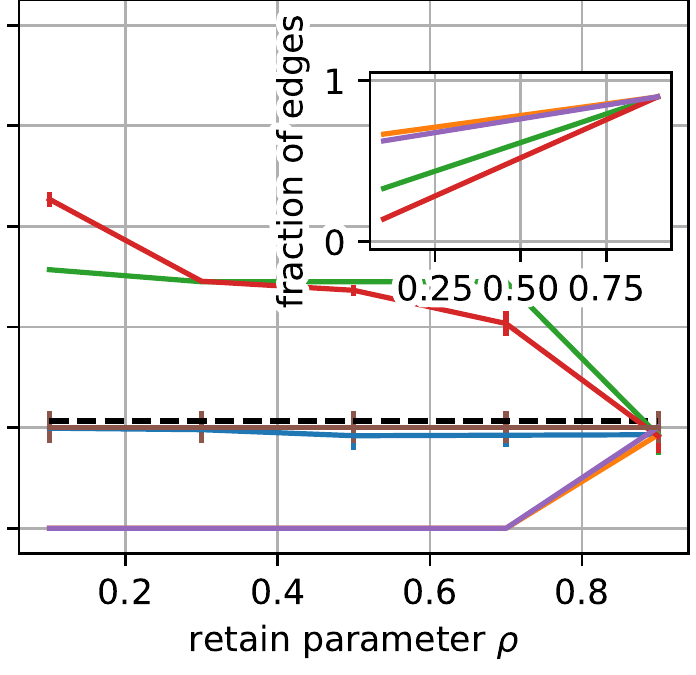}
			\caption{dblp \tiny{(moderately homophilic)}}
			\label{fig:dblp}
		\end{subfigure}
		\begin{subfigure}[b]{0.386847\columnwidth}
			\includegraphics[width=\textwidth]{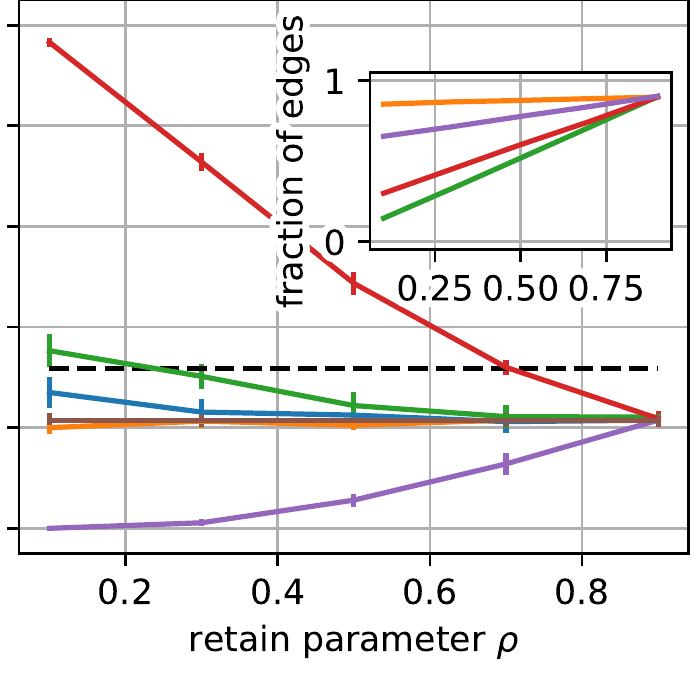}
			\caption{aps \tiny{(strongly homophilic)}}
		\end{subfigure}
	\end{center}
\caption{\textit{Effect of node attribute-based noise on real datasets}. {Similar to Fig.~\ref{fig:retain} we plot the influence of the retain parameter $\rho$ on networks with different strength of homophily. Networks are sorted by modularity of their label partition which can be  seen as a proxy for homophily, if the connectivity of the network is indeed driven by homophily. However, the networks here do not have a one-to-one correspondence with the synthetic networks regarding minority size and homophily.
Nonetheless we can see that similar effects are present in real and synthetic world networks, with dblp being the notable exception where some saturation happens already at around $\rho=0.7$}.}
\label{fig:real1}
\end{figure*}

\begin{figure*}[hbt]
	\begin{center}
	\begin{subfigure}[c]{2.0\columnwidth}
				\includegraphics[width=\textwidth]{figures/legend_struct10.pdf}
		\end{subfigure}
		\begin{subfigure}[b]{0.452612\columnwidth}
			\includegraphics[width=\textwidth]{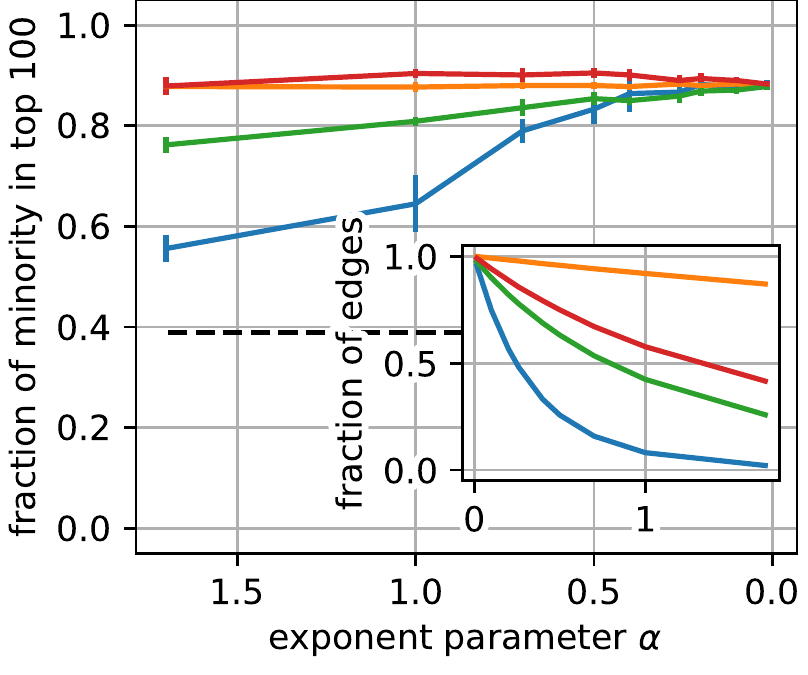}
			\caption{brazil\tiny{(strongly heterophilic)}}
		\end{subfigure}
		\begin{subfigure}[b]{0.386847\columnwidth}
			\includegraphics[width=\textwidth]{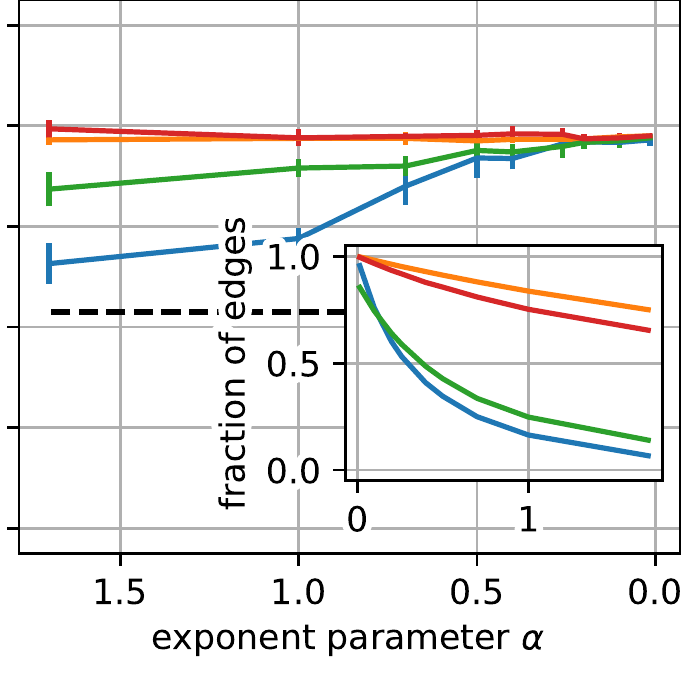}
			\caption{pok\tiny{(strongly heterophilic)}}
		\end{subfigure}
		\begin{subfigure}[b]{0.386847\columnwidth}
			\includegraphics[width=\textwidth]{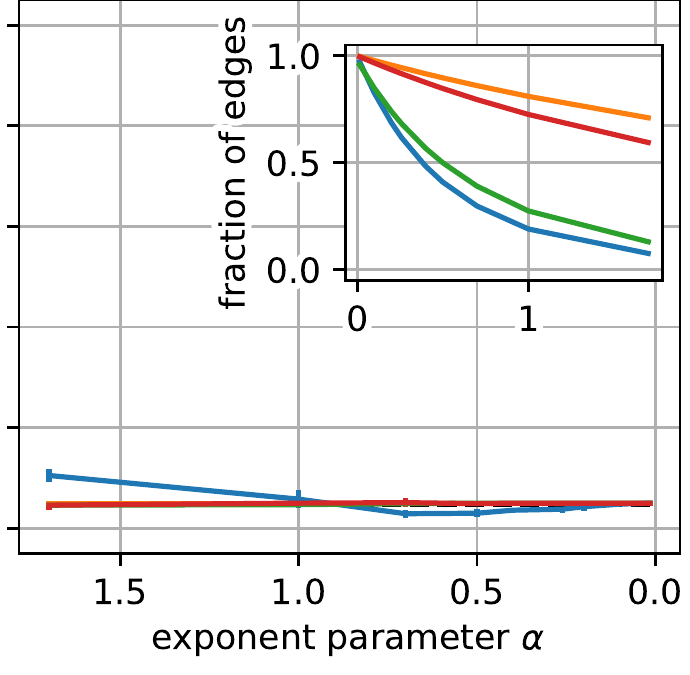}
			\caption{github\tiny{(neutral)}}
		\end{subfigure}
		\begin{subfigure}[b]{0.386847\columnwidth}
			\includegraphics[width=\textwidth]{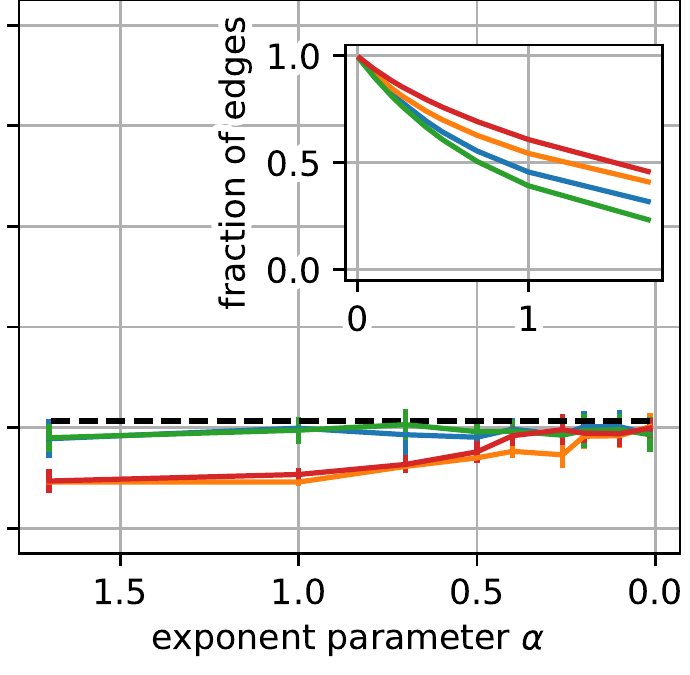}
			\caption{dblp\tiny{(moderately homophilic)}}
		\end{subfigure}
		\begin{subfigure}[b]{0.386847\columnwidth}
			\includegraphics[width=\textwidth]{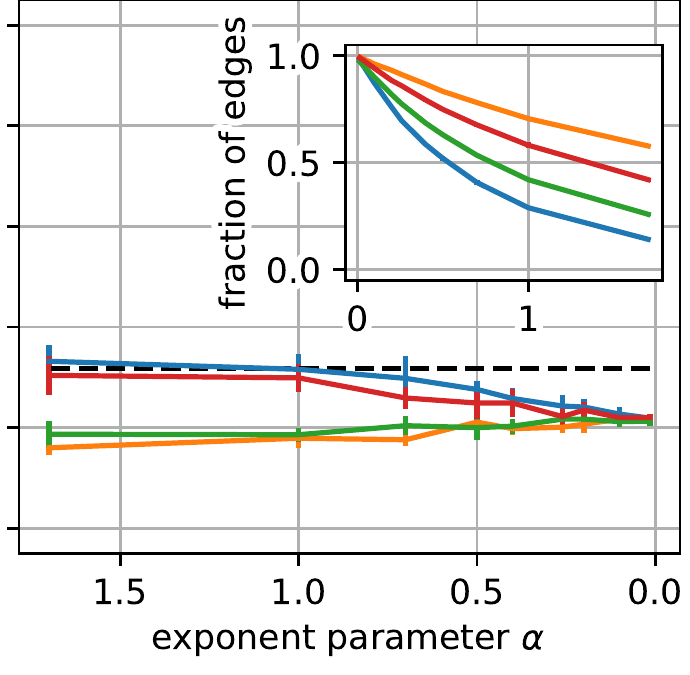}
			\caption{aps\tiny{(strongly homophilic)}}
		\end{subfigure}
	\end{center}
\caption{\textit{Effect of structure-based noise on real data sets.} In analogy to Fig.~\ref{fig:retain2} we show the impact of structure-based noise. 
We  see that in the heterophilic regime the effects of noise are similar to those in synthetic networks. 
In the homophilic regime different effects can be observed. 
For instance, for dblp removing edges within the strong community structure with inverse Jaccard noise leads to a disadvantage of the minority. 
Overall we can see that structure-based noise can also significantly affect real world networks. This effect can sometimes be even more prominent or different than in synthetic BA networks.}
\label{fig:real2}
\end{figure*}

\section{Experiments on real-world networks}
\label{sec:experiments_realworld}
In this section, we investigate whether the observations made for synthetic datasets hold with empirically observed networks. Even if these data are most likely not error-free to start with, they provide us with real topologies of complex networks that may not be captured by a synthetic model.
For these empirically observed networks, we can also test whether systematic noise based on our model would lead to significant effects. 
Using real-world networks in this way, we follow standard practice in the literature of assessing the effects of non-systematic errors \cite{murai_estimating_2019}.


\subsection{Real world networks}

We assess the impact of noise on a wide range of social networks with available binary attributes, see Table \ref{tab:real_data} for an overview. We report assortativity and modularity for all networks as a proxy for the homophily parameter. 
Note again the difference between node attribute- and structure-based groups: In the case of the synthetic networks, the correspondence was controlled by the generation process. This is different in the case of empirical networks. We therefore have to note that the modularity parameter of the label partition is only a proxy for homophily. 

For the datasets brazil, github and dblp the labels correspond to gender\footnote{We consider gender labels to be binary here due to data availability, being aware of the deficit of not explicitly considering non-binary gender identities.} 
For the pusokram (POK) network the labels were previously inferred using a maxcut approach, based on the assumption of a predominance of heterosexual relationships. We can therefore only label the two groups as majority/minority, as in \cite{karimi_homophily_2018}. 
For the APS network, labels reflect scientific sub-fields. 

\subsection{Results on empirical networks}
Figures \ref{fig:real1} and \ref{fig:real2} summarize the effects of applying different noise types to the empirical networks. Overall, we observe similar effects as for the synthetic networks when we consider them in order of their modularity of the label partition. 
This suggests that key factors for the implications of edge uncertainty on node centrality are already captured well by the comparatively simple modified BA model. 

However, there are some notable exceptions: In the dblp co-authorship network, a saturation already sets in for a retain parameter of around $\rho =0.7$. Beyond that value, there are no more significant effects to be observed when dropping more edges, independent of the noise type.
As another main discrepancy, the difference between Jaccard-based and centrality-based noise types with respect to the number of dropped edges is much weaker in empirical  to synthetic networks. 
 
This might be due to the fact, that centrality-noise mostly affects low-low degree connections. 
Such low-low degree connections occur more often in the empirical datasets due to local clustering, which is absent in Barabasi-Albert-based models.

The results of our simulations on real-world networks thus lead to similar results as the analysis of synthetic BA-networks. This suggests that the homophily of the network is responsible for some of the effects of systematic bias on minority representation in degree-based rankings.
\begin{table}[bht]
\caption{\textit{Overview of the empirical datasets}.{We use $Q_{mod}$/$A$ to denote the modularity/assortativity of the partition based on minority and majority label. The modularity is not obtained from an optimisation. Both metrics may be interpreted as a proxy for the homophily in the dataset. The fraction of the minority in the dataset is denoted by $f$. Our datasets cover a wide range of sizes, minority fraction and modularity/assortativity.}
    }
\small 
    \centering
    \begin{tabular}{l | r rrrr r}
     name & topic  &   nodes &   edges &  $Q_{mod}$   & $f$  & $A$\\\hline
      brazil \cite{rocha2010information} & sexworker &   15k &   38k &   -0.500  & 39\% & -1.00\\
        pok \cite{holme2004structure} & online dating&   25k &   25k &  -0.349   & 43\% & -0.84\\
      github \cite{github}                        &followers&  119k &  248k & 0.004     & 5\%  & -0.03\\
        dblp \cite{karimi2016inferring}           &coauthorship&  185k &  619k &     0.027 & 21\% & -0.15\\
         aps \cite{github}                      & citation&   1k &    3k &  0.346    & 32\% &  0.74\\
    \end{tabular}
    \label{tab:real_data}
\end{table}

\section{Discussion}
\label{sec:Discussion}\label{ssec:additional_considerations}
The main result of our work is that systematic errors can give rise to significant effects on subsequent network analysis. This emphasises that researchers should account for the possibility of such errors more carefully, e.g., by checking certain bias hypotheses using our framework, and including knowledge on systematic edge errors into their analysis, where possible.
Interestingly, we observed similar results when investigating the effects of systematic edge errors in real-world and synthetic networks generated by an augmented Barabasi-Albert model.
This may suggest that, at least for degree-based rankings, network properties (like clustering) are less important for the representation of minorities in rankings under systematic noise. 
These aspects should be investigated in future work.

Clearly, our study is only a first step towards better understanding of the effects of systematic edge errors on network analysis. 
For instance, effects that arise from the addition of edges instead of deletion are left to be investigated in future work. 
However, our error model provides a platform to account for a wide range of systematic errors that are in need of better understanding, and will thus hopefully trigger further research in this direction.
We emphasize that our model is not restricted to binary node attributes, it allows for continuous attributes such as age to simulate e.g. age discrimination. 

Moreover, the ideas presented here can be easily extended to directed and/or weighted networks.
In this case we could also consider asymmetric biases, e.g. of groups reporting differently on each other (in the context of directed graphs),  or edge-weight thresh-holding effects on ties (in the context of weighted graphs).
Such problems are for example present in Wikipedia Clickstream data, where the number of occurrences of each pairs of page requests and referers is only counted if exceeding 10 requests \cite{rodi_search_2017}.
Eventually, we may consider correlated edge errors, which can arise, for instance, if nodes selectively do not report a whole set of connections of a particular type, or replicate the reporting behavior of other nodes.
More broadly, we may also want to include effects that arise from sampling nodes, or having uncertain or incomplete node (attribute) information.

In our experiments we considered synthetic and empirical networks and removed edges based on different systematic errors. 
We thus treated the initial networks effectively as an accurate system representation, even though we can only assert this for the synthetically generated networks.
However, the empirical networks used are potentially not error free themselves.
Yet, such network often serve as a (more or less accurate) proxy for real network structure. 
In this context, it should also be acknowledged that multiple types of biases may lead to the same kind of observation errors.
For instance, in a network, whose structure may be due to homophilic interactions between socially alike groups, reporting bias correlated with the group structure (modelled by node attribute-based edge noise) may lead to similar effects on the network than noise arising due to an availability bias correlated with a strong neighborhood overlap (modelled by a form of structure-based noise).
While the resulting edge noise may thus be the same, the interpretation of its cause can be very different. 
This underlines that without additional information it is impossible to infer the  reason for a specific edge structure from a single network observation, and we can only use models to investigate potential implications. 

In this work we study the impact of systematic edge errors motivated by social biases.
Although we provide detailed motivation for the semantics of such edge error in form of certain biases, we cannot ascertain that an empirically observed network has been subject to a particular systematic edge error (without any additional information). 
This is somewhat reminiscent of the correlation vs. causation debate when considering homophily and contagion in observational social network studies, as investigated in \cite{shalizi_homophily_2011}. 

Finally, we have focused on degree-based rankings within the scope of this paper. 
However, this is obviously not the only network analysis task that can be affected by noise. 
In future work, we plan to investigate the effects of systematic errors on other types of centrality such as eigenvector centrality, or entirely different analysis methods such as community detection.

\subsection*{Conclusion}
We introduced a general framework for simulating systematic edge errors in attributed networks. 
Our framework discriminates between node attribute-based errors, such as label congruence-based and label specific errors, as well as  network structure-based errors, such as neighborhood overlap-based errors and centrality based-errors.
We applied this simulation framework to investigate the representation of minorities in rankings based on the degree centrality of nodes with binary labels, representing two groups.
In our numerical simulations on synthetic networks we find that the effect of the systematic bias is dependant on the network topology: In heterophilic networks, \emph{majority/minority representations in rankings are sensitive to the type of edge error present}. 
In contrast, in homophilic networks we find that \emph{minorities are at a disadvantage regardless of the type of error present}. We also performed error simulations on real world networks, which led to similar effects as the synthetic networks. 
Our results emphasize that systematic errors can heavily influence results of network analyses and the nature of the effect depends on the specific data set. This emphasises the necessity of a flexible framework such as ours to enable researchers to account for systematic errors in their social network analysis. 

\section*{List of abbreviations}
\begin{itemize}
    \item BA model: Barabasi-Albert model
    \item POK: Pusokram network (network of online dating) \cite{holme2004structure}
    \item brazil: Network of sexworkers \cite{rocha2010information}
    \item github: Network of github follower \cite{github}
    \item dblp: Co-authorship network (dblp) \cite{karimi2016inferring} 
    \item aps: Citation network (American Physical Society) \cite{github}  
\end{itemize}

\section*{Declarations}
\subsection*{Availability of data and materials}
The datasets dblp, github and aps datasets are available at \url{https://github.com/frbkrm/NtwPerceptionBias/tree/master/datasets}.
The datasets brazil and pok are not puclicly available, you may contact the authors of corresponding publications in case you are interested in working with those.

The code used is publicly available at \url{www.github.com/Feelx234/unnet}.

\subsection*{Competing Interests}
The authors declare no competing interests.

\subsection*{Funding}
This work has been funded by the Federal Ministry of Education and Research (BMBF) and the Ministry of Culture and Science of the German State of North Rhine-Westphalia (MKW) under the Excellence Strategy of the Federal Government and the Länder. 
Michael T. Schaub and Leonie Neuhäuser acknowledge funding by the Ministry of Culture and Science (MKW) of the German State of North Rhine-Westphalia (“NRW Rückkehrprogramm”).

\subsection*{Authors' contributions}
All authors conceived and designed the study and wrote the paper.
F.S. and L.N. performed the numerical simulations and analyzed the datasets. 

\begin{acks}
We thank Fariba Karimi and Claudia Wagner for valuable discussions and comments.
\end{acks}

\urlstyle{same}
\bibliographystyle{ACM-Reference-Format}
\bibliography{bib}
\clearpage
\newpage
\appendix
\label{appendix}

\section{Sensitivity analysis of $k$}
\label{sensitivity}

\begin{figure}[b]
	\begin{subfigure}[c]{1.0\columnwidth}
			\includegraphics[width=\textwidth]{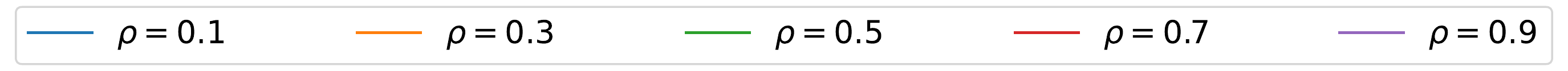}
		\end{subfigure}
	\begin{subfigure}[b]{0.49\columnwidth}
		\includegraphics[width=\textwidth]{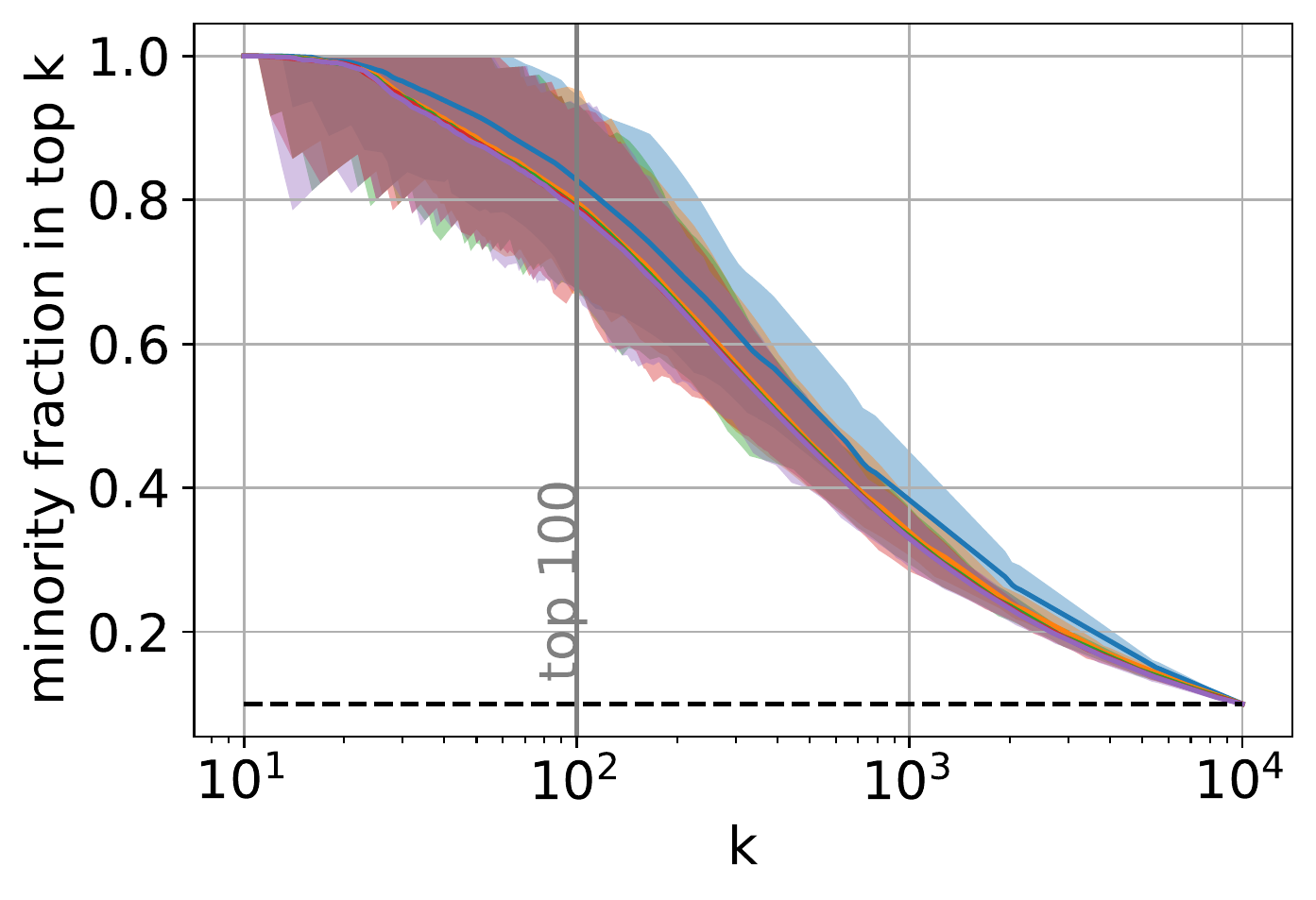}
		\caption{h=0.25 (heterophilic)}
	\end{subfigure}
	\begin{subfigure}[b]{0.49\columnwidth}
		\includegraphics[width=\textwidth]{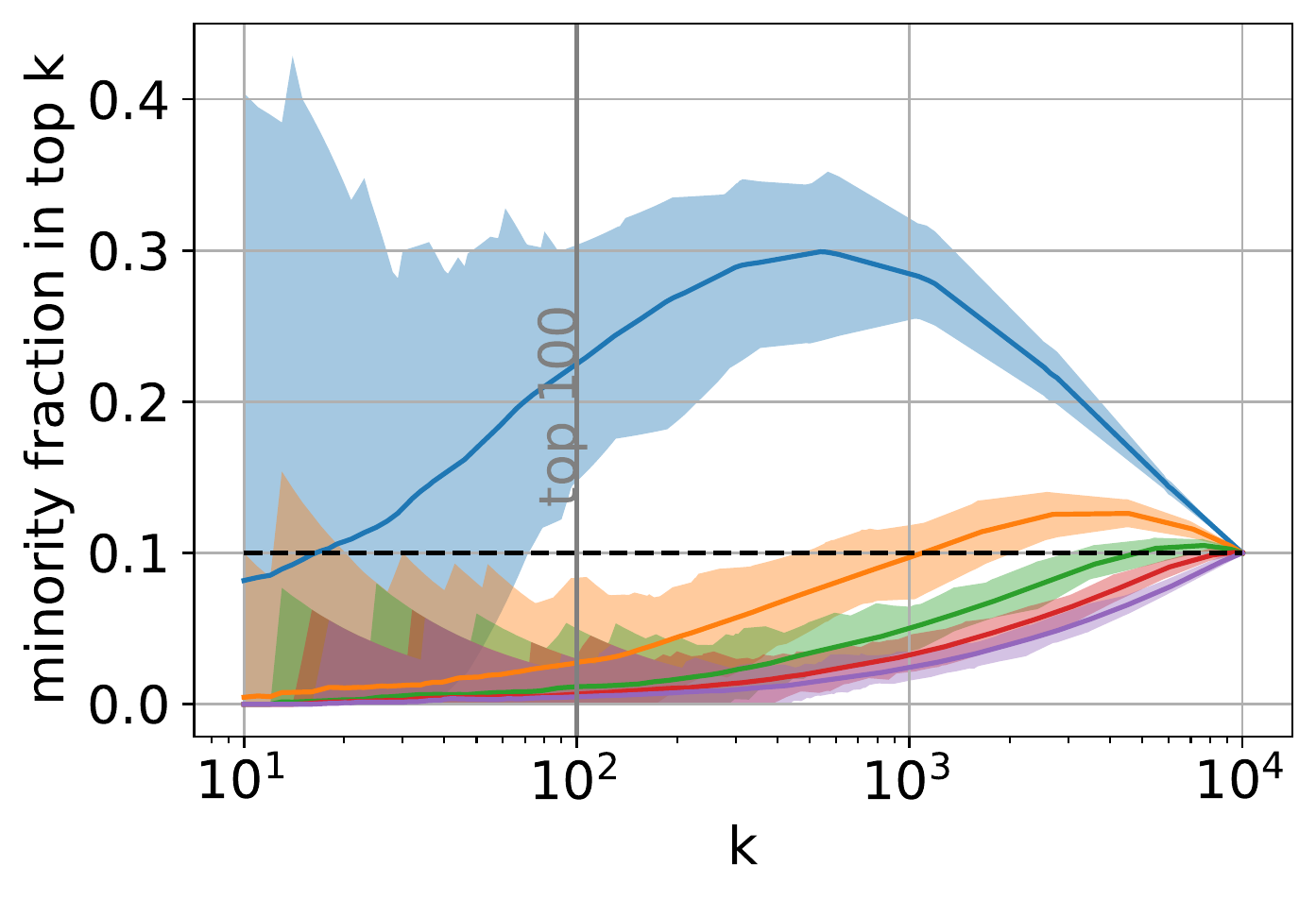}
		\caption{h=0.75 (homophilic)}
	\end{subfigure}
    \caption{\textit{Sensitivity of our results with regard to the choice of k.} {
    We visualize the fraction of the minority in the top k ranked nodes as a function of k, for different retain rates for majority associated noise.
    The solid line is the mean while the enveloping curve represent the range of values obtained in 10 iterations.
    Ties are resolved randomly. Due to large noise for the top ranked position we only start visualising for k > 10.
    We see that the effects displayed in Fig.~\ref{fig:retain} for k=100 are representative for a wider range of k values. }
    }
    \label{fig:sensitivity}
\end{figure}

In the main article, we investigated the representation of the minority in the top $k=100$ ranked nodes. 
We now examine the dependency of these results on $k$. 
Fig.~\ref{fig:sensitivity} shows the fraction of minority in the top $k$ ranked nodes in a heterophilic ($h=0.25$) and a homophilic ($h=0.75$) regime for different strengths of majority-noise. 
We observe that in the heterophilic regime, the over-representation of the minority in the top-k nodes is essentially independent of the parameter $k$ used. By contrast, in the homophilic regime we see a slight increase in representation with larger values of $k$ for values of $\rho > 0.1$. For $\rho=0.1$ the over-representation is present almost throughout the range of $k$ values.

\begin{figure}[b]
	\begin{subfigure}[c]{1.0\columnwidth}
		\includegraphics[width=\textwidth]{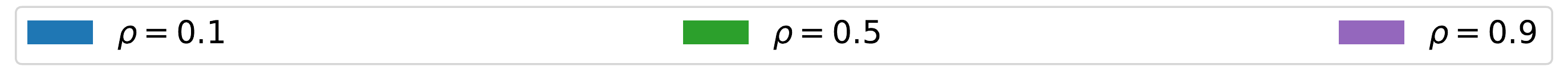}
	\end{subfigure}
	\begin{subfigure}[b]{0.49\columnwidth}
		\includegraphics[width=\textwidth]{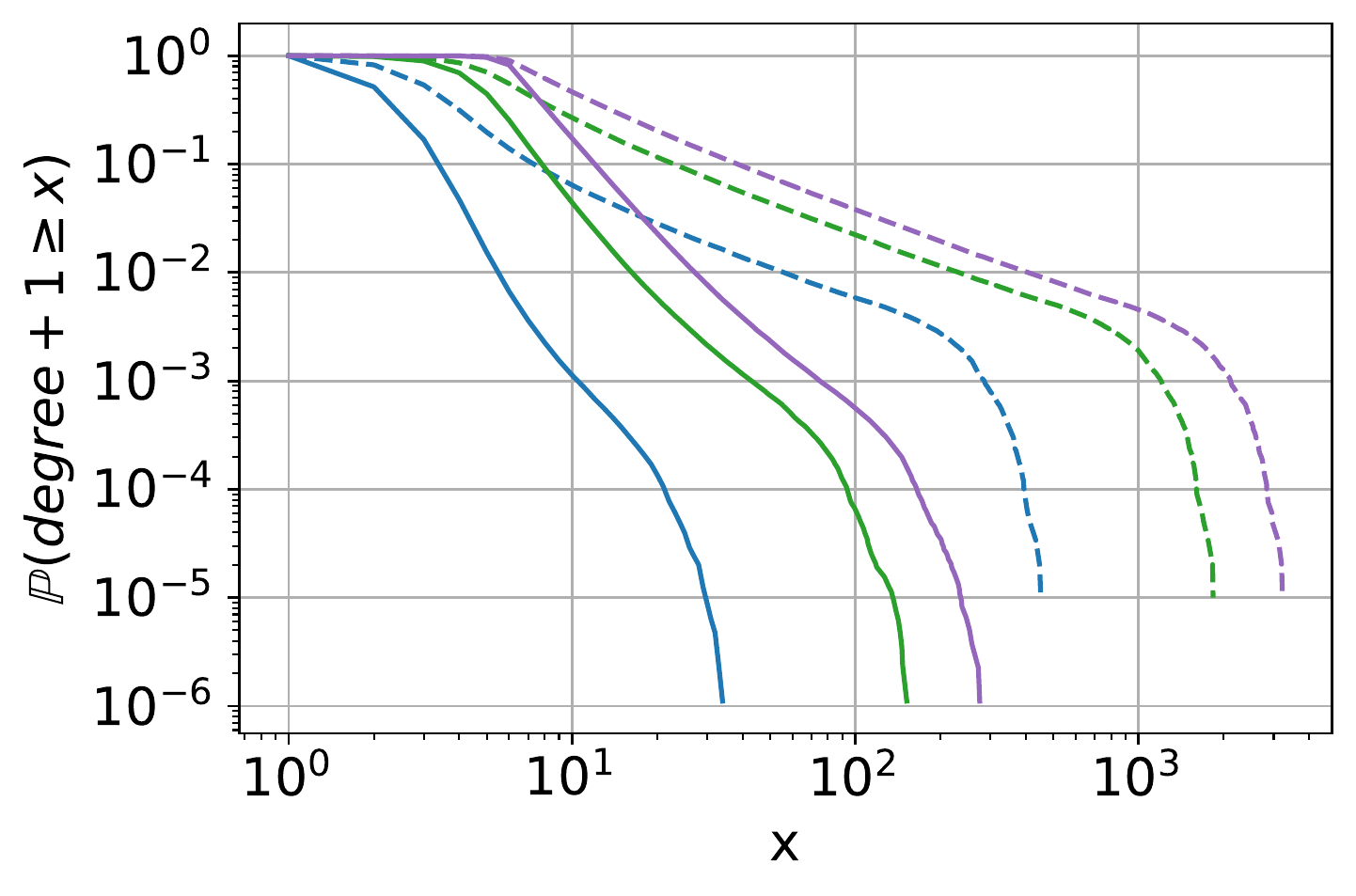}
		\caption{h=0.25 (heterophilic)}
		\label{fig:cummulative}
	\end{subfigure}
	\begin{subfigure}[b]{0.49\columnwidth}
		\includegraphics[width=\textwidth]{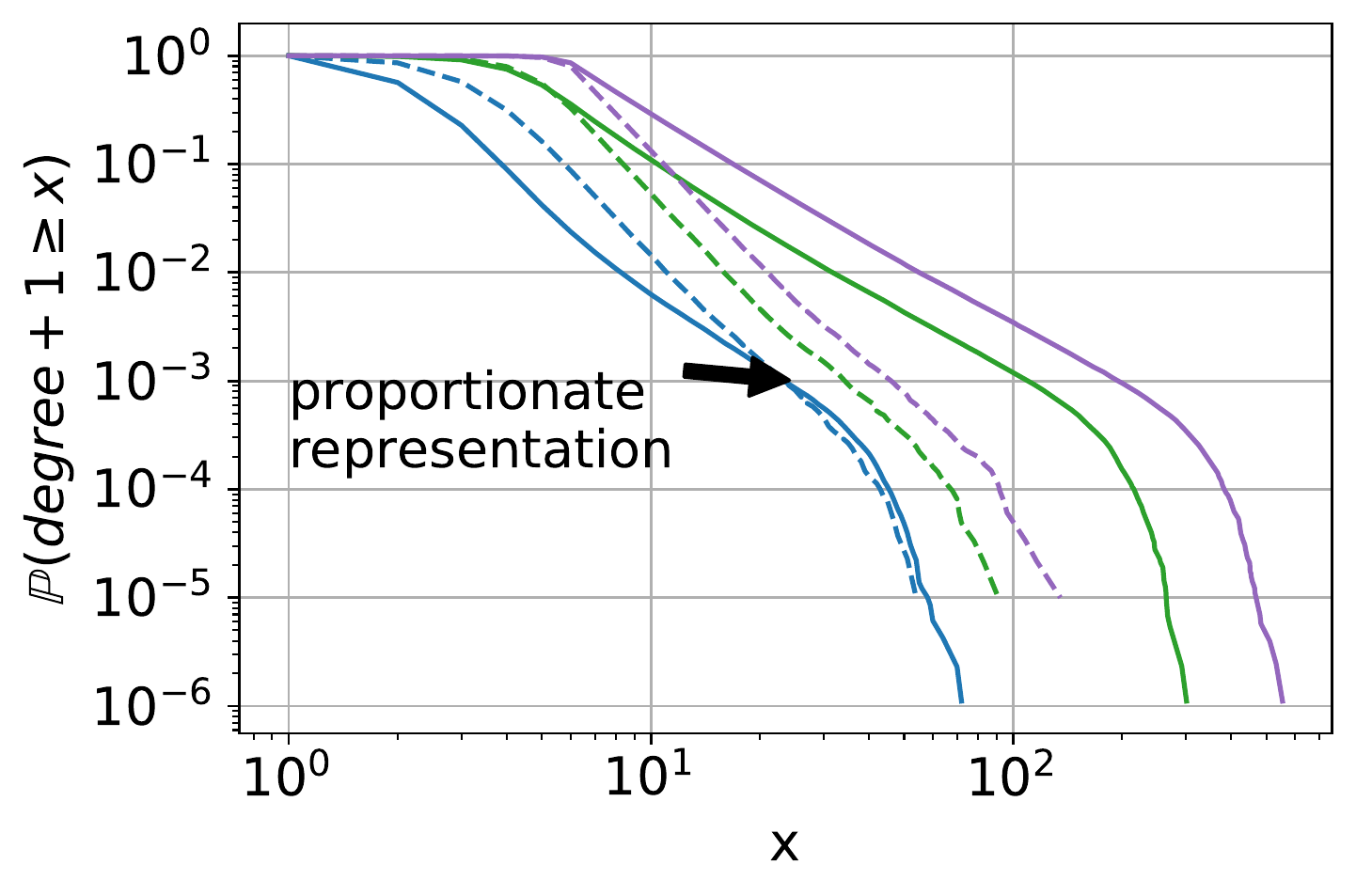}
		\caption{h=0.75 (homophilic)}
	\end{subfigure}
    \caption{\textit{Complementary cumulative distribution by class for synthetic networks and majority associated noise.}
    The statistics for minority/majority are represented by dashed/solid lines respectively. 
    Different colours correspond to different values of the retain parameter $\rho$.
    To increase clarity, standard deviations are not shown.
    For fixed $\rho$, comparatively higher values of the solid line indicate an over representation of the majority in degree rankings, equal values proportional representation.
    As expected the majority noise has a strong effect on the majority but also a minor effect on the minority. In heterophilic regimes the minority remains disproportionately represented for any ranking. In homophilic regimes for $\rho=0.1$ we can have a ranking with proportionate representation if we consider rankings above the top 0.1\% of nodes.} 
    \label{fig:cumulative}
\end{figure}

An alternative way to access the impact on node rankings across different values of $k$ is shown in Figure~\ref{fig:cumulative}.
There we see the impact of majority noise on the group specific complementary cumulative degree distribution.
While these plots do not allow to access a ranking directly, we can easily extract rankings for nodes with degree greater $x$, where $x$ is chosen on the abscissa.
Thus, differences between the solid and the dashed line at a specific value on the x-axis indicate an under- or over-representation of the minority. Proportional representation in degree-based rankings is achieved whenever both lines intersect.

We have highlighted one such point which shows that rankings, which include fewer than the top 0.1\% of nodes have a proportionate representation. 


\end{document}